\documentclass[10pt,a4paper]{article}
\pdfoutput=1
\usepackage{jheppub}

\usepackage[T1]{fontenc}
\usepackage{amssymb,amsmath}
\usepackage{lmodern, verbatim}
\usepackage{multirow}
\usepackage{bigstrut}

\DeclareMathOperator{\R}{Re}
\DeclareMathOperator{\I}{Im}
\newcommand{\ee}{\mathrm{e}}

\newcommand{\Iprod}[2]{\langle {#1}, {#2} \rangle}
\newcommand{\cG}{G}
\newcommand{\cK}{\mathcal{K}}
\newcommand{\cF}{\mathcal{F}}
\newcommand{\cbG}{\mathcal{G}}
\newcommand{\cL}{\mathcal{L}}
\newcommand{\cE}{\mathcal{E}}
\newcommand{\cV}{\mathcal{V}}
\newcommand{\Ww}{\mathcal{W}}
\def\cH{{\mathcal H}}
\newcommand{\im}{\mathrm{i}}
\newcommand{\nv}{n_\mathrm{v}}
\newcommand{\nh}{n_\mathrm{h}}
\newcommand{\cN}{\mathcal{N}}

\newcommand{\vsorb}{\mathcal{S}}

\def\slash#1{\rlap{\hbox{$\mskip 1 mu /$}}#1}      
\usepackage{array}
\usepackage{hhline}

\makeatletter
\newcommand{\thickhline}{%
    \noalign {\ifnum 0=`}\fi \hrule height 1.3pt
    \futurelet \reserved@a \@xhline
}
\newcolumntype{"}{@{\hskip\tabcolsep\vrule width 1pt\hskip\tabcolsep}}
\makeatother
\newlength{\arrayrulewidthOriginal}
\newcommand{\Cline}[2]{%
  \noalign{\global\setlength{\arrayrulewidthOriginal}{\arrayrulewidth}}%
  \noalign{\global\setlength{\arrayrulewidth}{#1}}\cline{#2}%
  \noalign{\global\setlength{\arrayrulewidth}{\arrayrulewidthOriginal}}}

\title{Ungauging black holes and hidden supercharges}
\author[a]{Kiril Hristov,}  \author[b]{Stefanos Katmadas,} \author[b]{and Valentina Pozzoli}
\affiliation[a]{Dipartimento di Fisica, Universit\'a di Milano-Bicocca,
I-20126 Milano, Italy}
\affiliation[b]{Centre de Physique Th\'eorique, \'Ecole Polytechnique, CNRS, 91128
Palaiseau, France}

\emailAdd{kiril.hristov [at] unimib.it} \emailAdd{stefanos.katmadas [at] cpht.polytechnique.fr} \emailAdd{valentina.pozzoli [at] cpht.polytechnique.fr}

\abstract{
We embed the general solution for non-BPS extremal asymptotically flat static and under-rotating black holes in abelian gauged $D\!=\!4$ $\cN\!=\!2$ supergravity, in the limit where the scalar potential vanishes but the gauging does not. Using this result, we show explicitly that some supersymmetries are preserved in the near horizon region of all the asymptotically flat solutions above, in the gauged theory. This reveals a deep relation between microscopic entropy counting of extremal black holes in Minkowski and BPS black holes in AdS. Finally, we discuss the relevance of this construction to the structure of asymptotically AdS$_4$ black holes, as well as the possibility of including hypermultiplets.
}

\preprint{CPHT-RR073.1012}

\begin{document}
 \maketitle

\section{Introduction}

The interplay between the macroscopic description of black holes in supergravity and
their corresponding microscopic description within string theory has been a source of important
insights into the structure of the theory. In this respect, the most detailed investigations
have been carried out for asymptotically flat black holes preserving some amount of supersymmetry, which provides additional control over various aspects of these systems. In particular, the microscopic counting of black hole entropy \cite{Strominger:1996sh, Maldacena:1997de} as well as the construction of
the corresponding black hole geometries \cite{Behrndt:1997ny, Denef:2000nb, Gauntlett:2002nw}, depend crucially on the presence of unbroken supercharges.

Beyond the supersymmetric sector, the non-BPS class of asymptotically flat black holes
in supergravity has attracted attention, based on a deeper understanding of the first
order systems underlying extremal static and under-rotating solutions (i.e.~rotating black holes
without an ergo-region) \cite{Khuri:1995xq, Rasheed:1995zv, Ortin:1996bz, Larsen:1999pp, LopesCardoso:2007ky, Gimon:2007mh, Goldstein:2008fq, Bossard:2009at, Bena:2009ev, Bossard:2009my, Bena:2009en, Bossard:2009bw, Bena:2009fi, Bossard:2011kz, Bossard:2012xs}. While these systems are
considerably more complicated than the corresponding BPS ones, they are in principle
exactly solvable, since they are described by first order differential equations. In
addition, various similarities to the BPS branch have been observed at the formal
level, for example through the existence of a fake superpotential
\cite{Ceresole:2007wx, Bellucci:2008sv, Ceresole:2009vp, Ceresole:2009iy, Bossard:2009we}.

More recently, the interest in four-dimensional black holes was extended to the more general case of asymptotically anti-de Sitter (AdS) spacetimes, described as solutions to gauged supergravity theories \cite{Caldarelli:1998hg,Sabra:1999ux,Cacciatori:2009iz,Dall'Agata:2010gj,Hristov:2010ri,Klemm:2012yg,Toldo:2012ec}. Although not fully exhaustive, the existing classification of black holes in AdS shows a rich variety of possibilities with both static and rotating BPS solutions, as well as new horizon topologies. While a microscopic account of their entropy is not available yet, they provide interesting new examples in the context of the AdS/CFT correspondence. In addition, understanding phase transitions of extremal and thermal black holes in this class could lead to insight into the phase structure of physically interesting field theories at strong coupling.

A priori, the above mentioned classes of black holes in Minkowski and AdS spaces are unrelated, as they usually arise as solutions to different supergravity theories and respectively in different string theory compactifications, when these exist. Consequently, the two systems are usually clearly distinguished and studied by different methods, while the problems of microscopic entropy counting for asymptotically flat and AdS black holes are viewed independently (\cite{Strominger:1996sh, Maldacena:1997de} vs.\ \cite{Berkooz:2006wc,Kinney:2005ej}). However, our purpose in this paper is to show that such distinction is not always present. In particular, we show that one can embed asymptotically flat non-BPS black hole solutions in certain special $D\!=\!4$ $\cN\!=\!2$ gauged supergravity theories. Moreover, we show that the attractor geometries, and therefore also the microscopic counting, for BPS black holes in AdS$_4$ \cite{Bellucci:2008cb,Cacciatori:2009iz,Dall'Agata:2010gj,Hristov:2010ri} and asymptotically 
flat extremal non-
BPS black holes \cite{Kallosh:2006ib, Tripathy:2005qp, Goldstein:2005hq, Sen:2005wa, Nampuri:2007gv, Ferrara:2007tu} fall within a common class of supersymmetric AdS$_2 \times$S$^2$ spaces\footnote{This class of horizons was called magnetic AdS$_2 \times$S$^2$ in \cite{Hristov:2012bk} for the reason that, just like for asymptotically magnetic AdS$_4$ spacetimes, the fermions flip their spin and become $SU(2)$ scalars \cite{deWit:2011gk,Hristov:2011ye}. We elaborate on this more in the following sections.}, or their rotating generalizations.

\begin{figure}[ht]
  \centering
 \includegraphics[scale=0.45]{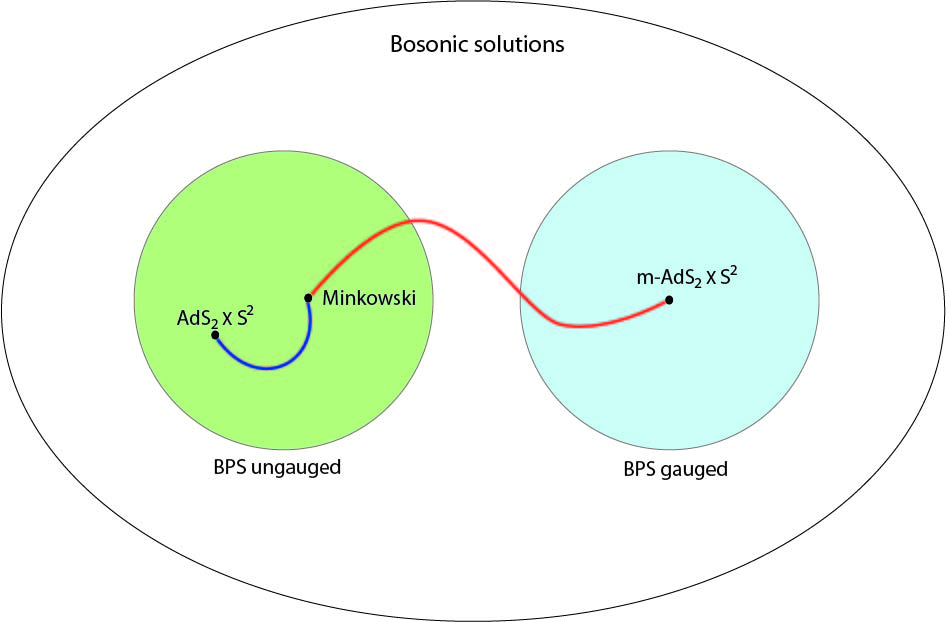} 
  \caption{\sf\small The two bubbles above represent the space of BPS solutions to ungauged and abelian gauged supergravity with a flat potential, as subspaces of all bosonic solutions, common to both theories. Note the presence of two distinct AdS$_2 \times$S$^2$ backgrounds that are supersymmetric only within one theory. The blue line represents the BPS black hole solutions, interpolating between Minkowski space and the fully BPS AdS$_2 \times$S$^2$. The solutions described in this paper, represented by a red line, interpolate between Minkowski and the so called magnetic AdS$_2 \times$S$^2$ vacuum.} 
  \label{fig:gaugedsols}
\end{figure}

Let us be slightly more precise and consider the bosonic Lagrangian of abelian gauged $\cN\!=\!2$ supergravity in four dimensions with an arbitrary number $\nv$ of vector multiplets and no hypermultiplets (i.e.\ we consider constant gauging Fayet-Iliopoulos (FI) parameters). Such theories are described in \cite{deWit:1984pk, deWit:1984px} and we give more details in the following sections. For presenting the main argument we only need to know that the bosonic part of the Lagrangian is modified with respect to the one for the ungauged theory, 
$ \cL_{0}^{bos}$, by the introduction of a scalar potential term for the vector multiplet complex scalars, $t^i$, $i=1\dots\nv$ as
\begin{equation}
	\cL_{g}^{bos} = \cL_{0}^{bos} + V(t,\bar{t})\ ,
\end{equation}
 with
 \begin{equation}
	V(t,\bar{t}) = Z_i(\cG)\,\bar Z^i(\cG) -3\,\left|Z(\cG)\right|^2\ ,
\end{equation}
where $G = \{g^I, g_I\}$ is a symplectic vector of arbitrary constant FI parameters and $Z(\cG)$, $Z_i(\cG)$ denote its scalar dependent central charges. An interesting possibility arises in broad classes of vector multiplet moduli spaces when the FI parameters are chosen in a way as to make the scalar potential identically zero \cite{Cremmer:1984hj}, without reducing the theory to the ungauged one. This requires at least one FI parameter to be non zero, thus leading to a different supersymmetric completion of the same bosonic Lagrangian, since $\cL_{g}^{bos} = \cL_{0}^{bos}$ when the potential $V(t,\bar{t})$ vanishes, but the fermionic sector of the gauged theory still involves the vector $\cG$ linearly. It is then immediately obvious that all purely bosonic background solutions of the ungauged supergravity are also solutions of this "flat" gauged supergravity. However, due to their different fermionic sectors, the supersymmetric vacua of the two theories do not coincide. It is in fact easy to show that
none of the BPS solutions of the ungauged theory are supersymmetric with respect to the gauged theory and vice versa (see section \ref{sec:BPShor}). This situation is summarised in Figure \ref{fig:gaugedsols}.

Given the above, it is not surprising that some known non-BPS solutions in ungauged supergravity might be supersymmetric in these special gauged theories. Indeed, our analysis shows that all extremal under-rotating black holes\footnote{In what follows, we refer for simplicity to under-rotating solutions having in mind that this includes also the static case, when the rotation vanishes.} preserve some supersymmetry in their near horizon region. Restricting to the static solutions, we further show that these horizon solutions are part of a larger class of supersymmetric horizons based on FI terms, that do not a priori satisfy the flat potential restriction and pertain to the static BPS black holes in AdS$_4$, \cite{Bellucci:2008cb,Cacciatori:2009iz, Dall'Agata:2010gj,Hristov:2010ri, deWit:2011gk}. It follows that one needs to address together the problems of microscopic entropy counting of asymptotically flat and AdS black holes in this case. We come back to this point in the concluding section of this paper, 
which we leave for more general discussion.

The following main sections of the paper address various aspects of the connection between solutions in gauged and ungauged theories sketched above, and are largely independent of each other. For the convenience of the reader, we give an overview of the main results presented in detail in each of these sections, as follows.

In section \ref{sec:static-ungauging}, we show that asymptotically flat extremal non-BPS black holes can be viewed as solutions to $\cN\!=\!2$ abelian gauged supergravity, if the FI gaugings are assumed to be such that the potential is trivial. The prime example of gauged theories with an identically flat potential can be found within the interesting class of cubic prepotentials arising in the ungauged case from Calabi-Yau compactifications of string/M-theories, as first discussed in \cite{Cremmer:1984hj}. This condition is enforced by introducing a Lagrange multiplier, which allows us to write the action of the extended system as a sum of squares, similar to the 1/4-BPS squaring in \cite{Dall'Agata:2010gj}, while demanding that the metric is asymptotically flat, as is appropriate for a theory without a potential. The result is a first order system that is otherwise identical to the corresponding one describing
asymptotically AdS$_4$ BPS solutions, except for the presence of the Lagrange multiplier,
which is determined independently by its own equation of motion. We finally show that
the general non-BPS solutions, in the form cast in \cite{Bossard:2012xs}, are solutions to the system above,
once a suitable regularity constraint is imposed. This includes the identification of the
auxiliary very small vector appearing in that work as the vector of FI terms in the gauged
theory.

In section \ref{sec:BPShor}, we consider the near horizon limit of our system, making use of the fact
that the Lagrange multiplier above reduces to an irrelevant constant. It follows that
the attractor equations for general asymptotically flat static black holes
can be cast as a particular case of the attractor equations of
gauged supergravity \cite{Dall'Agata:2010gj}. The latter are expected to belong to the
family of attractors in \cite{deWit:2011gk} preserving four supersymmetries, which
we show explicitly to be the case. We therefore obtain the result that all static
non-BPS attractors in ungauged supergravity can be viewed as $1/2$-BPS attractors
once embedded in an abelian gauged supergravity with appropriately tuned FI terms.
Finally, we generalise this result in section \ref{sec-rot-attr}, where we show that
the under-rotating attractors of all asymptotically flat black holes \cite{Astefanesei:2006dd},
again in the form described in \cite{Bossard:2012xs}, preserve two supercharges,
i.e.~they are $1/4$-BPS. Note that this implies the presence of the same
number of supercharges in the near horizon region of any particular center of
a non-BPS multi-center solution.

In section \ref{sec:AdS} we consider the 1/4-BPS flow equations of \cite{Dall'Agata:2010gj}
for gauged supergravity, without restricting the FI terms, and show that some of the
structures found in asymptotically flat solutions are present in the more general case.
Most importantly, the regularity constraint used to define the single center
flow in \cite{Bossard:2012xs} is shown to hold even for an unrestricted vector of FI
terms. Since this constraint implies that only half of the charges can be present
once a vector of gaugings is specified, we expect it to be of importance in understanding
the moduli space of AdS$_4$ solutions. In section \ref{sec:hypers} we briefly discuss
the possibility of further embedding the asymptotically flat solutions above in
theories with gauged hypermultiplets. In such a scenario, the additional potential
induced by the hypermultiplets must also vanish, which we show to be possible in
rather generic theories that result from string compactifications.

We conclude in section \ref{sec:conc}, where we comment on the implications of our
results for microscopic models of black holes and on relations to recent developments
in the construction of non-BPS supergravity solutions. Finally, in the appendices we present some details
of our conventions, we extend the discussion of section \ref{sec:static-ungauging} to
the embedding of asymptotically flat under-rotating solutions in gauged supergravity,
and we discuss an example solution in some detail for clarity.

\section{Ungauging black holes}
\label{sec:static-ungauging}

In this section, we present the essential argument of the ungauging procedure for black hole solutions and
provide an explicit example by considering the static case for simplicity.
The starting point is the bosonic action for abelian gauged supergravity
\cite{deWit:1984pk, deWit:1984px}, which reads
\begin{equation}
\label{Ssugra4D}
S_\text{4D}=\frac{1}{16\pi}\int_{M_4} \Bigl(R\star 1 - 2\,g_{i\bar\jmath}\,d t^i\wedge\star d \bar{t}^{\bar\jmath} - \tfrac12 F^I\wedge G_I + 2\,V_g\,\star 1\Bigr),
\end{equation}
and describes  neutral complex scalars $t^i$ (belonging to the $\nv$ vector multiplets) and abelian gauge fields $F_{\mu\nu}{}^I$, $I={0,\,i}=0,\dots \nv$ (from both the gravity multiplet and the vector multiplets), all coupled to gravity\footnote{We refer to appendix \ref{app:conv} for some of our conventions in $\cN\!=\!2$ supergravity.}. The dual gauge fields  $G_{\mu\nu}{}_I$ are given in terms of the field strengths and the scalar dependent period matrix $\cN_{IJ}$, by
 \begin{equation}\label{G-def}
  G^-_{\mu\nu}{}_I = \cN_{IJ} F^-_{\mu\nu}{}^J\,,
 \end{equation}
where the expression for the period matrix will not be needed explicitly. Finally, the scalar potential $V_g$ takes the form
\begin{equation}\label{gau-pot}
V_g= Z_i(\cG)\,\bar Z^i(\cG) -3\,\left|Z(\cG)\right|^2 = \Iprod{\cG}{\mathrm{J}\,\cG} -4 \left|Z(\cG)\right|^2 \,,
\end{equation}
where we used the definition of the scalar dependent matrix $\mathrm{J}$ in \eqref{VBH-def}, and the symplectic vector $\cG=\{g^I, g_I\}$ stands for the FI terms, which control the coupling of the vector fields. In the abelian class of gaugings we consider in this paper, these couplings occur only in the fermionic sector of the theory, through the minimal coupling of the gravitini to the gauge fields, as the kinetic term is proportional to
\begin{gather}
\epsilon^{\mu\nu\rho\sigma} \bar\psi_{\mu}{}_i\gamma_\nu\,D_\rho\psi_{\sigma}{}^i \equiv
\epsilon^{\mu\nu\rho\sigma}
\bar\psi_{\mu}{}_i\gamma_\nu\left( \partial_\rho + \tfrac{i}2\,\Iprod{G}{A_\rho} \right)\psi_{\sigma}{}^i\,,
\\
\Iprod{G}{A_\mu}=g_I A_\mu{}^I - g^I A_{\mu}{}_I\,.\nonumber
\end{gather}
This coupling is in general non-local, due to the presence of the dual gauge fields $A_{\mu}{}_I$. However, as for any vector, $G$ can always be rotated to a frame such that it is purely electric, i.e.~$g^I=0$, leading to a local coupling of the gauge fields. More generally, one can consider couplings of magnetic vectors as well, using the embedding tensor formalism \cite{deWit:2005ub, deWit:2011gk}, which requires the introduction of extra auxiliary fields.

For the theories discussed in this paper however, the bosonic action is only affected through the nontrivial potential \eqref{gau-pot}, which can be straightforwardly written in an electric/magnetic covariant way, as above. Based on this observation, we take the pragmatic view\footnote{A similar point of view was used in \cite{Dall'Agata:2010gj}.} of using covariant versions of all quantities, keeping in mind that while the equations involving fermions strictly apply only to the electrically gauged theory, all results for the bosonic backgrounds must necessarily be covariant under electric/magnetic duality. We therefore employ covariant notation when dealing with the bosonic sector and covariantise the fermionic supersymmetry variations (see section \ref{sec:BPShor}), so that we do not have to choose a frame for the FI terms explicitly.

Given these definitions, we now discuss the connection of the gauged action above to the ungauged theory, at the bosonic level. As one would expect, ungauged supergravity is immediately recovered by putting $\cG=0$ in the above Lagrangian. However, it turns out that this is not the most general choice if one is interested in the bosonic sector only, as the scalar potential is not positive definite, and one can find nonzero $\cG$ for which the potential is identically zero \cite{Cremmer:1984hj}. The appropriate FI terms are then described by a so-called very small vector, characterised by
\begin{gather}\label{eq:doub-crit}
3\, |Z(\cG)|^2= Z_i(\cG)\,\bar Z^i(\cG) \,.
\end{gather}
In the context of symmetric scalar geometries such vectors are viewed as points of the doubly critical orbit, $\vsorb$, defined as the set of vectors such that \eqref{eq:doub-crit} is satisfied for any value of the scalars
\cite{Ceresole:2010nm, Borsten:2011ai}. Explicitly, they can be always brought to the frame where there is only one component, e.g.
\begin{equation}\label{small-exam}
\cG_0 = g\,\{0\,, \delta^{\,0}_{I} \}\,,
\end{equation}
but we will not impose any restriction other than \eqref{eq:doub-crit}. In what follows, we will be using the fact that this orbit exists for symmetric models, but the same arguments can be applied whenever \eqref{eq:doub-crit} has a solution for any model. For example, \eqref{small-exam} is an example solution for any cubic model, symmetric or not, and one may construct more general examples by acting with dualities\footnote{In fact, our treatment, as well as those of \cite{Dall'Agata:2010gj, Bossard:2012xs} whose results we connect, is duality covariant, so that the form of the prepotential is not fixed.}.

Given this special situation, it is natural to consider the possibility of finding asym\-pto\-ti\-ca\-lly flat backgrounds in a gauged theory with a flat gauging as above. Indeed, a vector of parameters in a doubly critical orbit was recently encountered in \cite{Galli:2010mg, Bossard:2012ge, Bossard:2012xs}, which considered the general under-rotating extremal black hole solutions in ungauged extended supergravity. As we now show, the presence of such a vector in asymptotically flat solutions can be seen to arise naturally by viewing the ungauged theory as a gauged theory with $G\in\vsorb$, leading to an interpretation of the auxiliary parameters introduced in \cite{Galli:2010mg, Bossard:2012ge, Bossard:2012xs} as residual FI terms.

\subsection{Squaring of the action}

In order to study extremal solutions in abelian gauged supergravity with a flat potential, we consider the squaring of the action for such backgrounds \cite{Denef:2000nb}, following closely the derivation of the known flow equations of \cite{Dall'Agata:2010gj} for asymptotically AdS$_4$ black holes that preserve $1/4$ of the supersymmetries. The only additional ingredient we require is the introduction of a Lagrange multiplier that ensures the flatness of the potential.

As we are interested in static solutions, we consider a spherically symmetric metric ansatz of the type
\begin{gather}\label{4dmetric}
  ds^2 = -\ee^{2U}d t^2  + \ee^{-2U} \left( dr^2 + \ee^{2\psi} d\theta^2 + \ee^{2\psi} \sin^2{\theta} d\phi^2 \right)\,,
\end{gather}
as well as an analogous ansatz for the gauge field strengths
\begin{equation}
F^{I}_{\theta \varphi}
=\tfrac{1}{2}\,p^{I} \,\sin \theta\,,
\qquad
G_{I}{}_{\theta \varphi}
=\tfrac{1}{2}\,q_{I}\, \sin \theta\,.
\end{equation}
Here, $\ee^{U}$, $\ee^{\psi}$ are two scalar functions describing the scale factor of the metric and the three dimensional base space, and $\Gamma=\{p^I\,, q_I\}$ denotes the vector of electric and magnetic charges.
Using these ansatze, the action \eqref{Ssugra4D} can be shown to be expressible in the form \cite{Dall'Agata:2010gj}
\begin{equation}\label{BPS2}
\begin{array}{rcl}
S_{1d}&=&\displaystyle\int dr\left\{-\frac12\ee^{2(U-\psi)}
\Iprod{\cal E}{\mathrm{J}\,{\cal E}} -\ee^{2\psi}\left[ (Q_r + \alpha') + 2 \ee^{-U} \, {\rm Re}(\ee^{-i \alpha} { W}) \right]^{2}\right.\\[5mm]
&&\displaystyle-\ee^{2\psi}\left[\psi'-2\ee^{-U}\, {\rm Im}(\ee^{-i \alpha} W) \right]^{2}-\left(1+\Iprod{G}{\Gamma}\right)\\[4mm]
&&\displaystyle\left.-2\frac{d\phantom{r}}{dr}\left[\ee^{2\psi-U}\,{\rm Im}(\ee^{-i \alpha} W) +\,\ee^{U}\,{\rm Re}(\ee^{-i \alpha}Z)\right]\right\}\,,
\end{array}
\end{equation}
where $\ee^{-i\alpha}$ is an arbitrary phase, we defined
\begin{equation}\label{E-def}
{\cal E} \equiv  2\ee^{2 \psi}\left(\ee^{-U}{\mathrm{Im}}(\ee^{-i\alpha}{\cal V})\right)'
-\ee^{2(\psi-U)}\, \mathrm{J}\,\cG
 +4\ee^{-2U}\,{\rm Re}(\ee^{-i \alpha} { W})\,{\mathrm{Re}}(\ee^{-i\alpha}{\cal V})
+\Gamma\,,
\end{equation}
and we introduced special notation for the central charges of $\Gamma$ and $G$ as
\begin{equation}\label{eq:W-def}
 Z\equiv Z(\Gamma)\,, \qquad  Z_i\equiv Z_i(\Gamma)\,,\qquad
 W\equiv Z(G)\,, \qquad  W_i\equiv Z_i(G)\,,
\end{equation}
for brevity.
The equations of motion following from this effective action imply the equations of motion for the scalars as well as the $tt$-component of the complete Einstein equation, whereas the remaining Einstein equations are identically satisfied upon imposing the Hamiltonian constraint
\begin{eqnarray}\label{eq:Ham-constraint}
\ee^{2\psi}\psi^{'}{}^2 -1 - \ee^{2\psi}U^{'}{}^2
- \ee^{2\psi} g_{i\bar \jmath}\,t^{i\, \prime}\,\bar t^{i\, \prime}
+ \ee^{2(U-\psi)} V_{\text{\tiny BH}} +\ee^{2(\psi-U)}V_g =0\ .
\end{eqnarray}
Solutions of this system have been discussed in \cite{Cacciatori:2009iz, Hristov:2010ri, Dall'Agata:2010gj}. These works analysed in some detail the asymptotically AdS$_4$ solutions associated to generic values of the gaugings $G$, and we return to this case in section \ref{sec:AdS}.

We now proceed to an analysis of the ungauged limit of the bosonic sector of the theory, by imposing that the vector of FI gaugings $G$ lies in the doubly critical orbit, $G\in\vsorb$, so that the potential is identically flat for any value of the scalars. Given the homogeneity of the potential in terms of $G$, we introduce a Lagrange multiplier in the original action \eqref{Ssugra4D}, through a rescaling of the type
\begin{equation}\label{static-rescal}
G\rightarrow \ee^{\varphi}G \,.
\end{equation}
Here and henceforth $\varphi$ will be treated as an independent field, whose equation of motion is exactly \eqref{eq:doub-crit}, enforcing the flatness of the potential. One can then write the action as a sum of squares in a similar way as above, up to an extra term originating from the partial integration involved. The result reads
\begin{align}\label{eq:stat-squaring}
S_{1d}=\int dr &\left\{ \, -\frac12 \ee^{2(U-\psi)} \Iprod{\cal E}{\mathrm{J}\,{\cal E}}
 -\ee^{2\psi}\left[ (Q_r + \alpha')
 + 2 \ee^{(\varphi-U)} \, {\rm Re}(\ee^{-i \alpha} { W}) \right]^{2}\right.\nonumber\\
& -\ee^{2\psi}\left[\psi'-2\ee^{(\varphi-U)}\, {\rm Im}(\ee^{-i \alpha} W) \right]^{2}
\nonumber\\
&\left. -\left(1+\ee^{\varphi}\,\Iprod{G}{\Gamma} \right)
 +2\,\ee^{2\psi}\,\ee^{(\varphi-U)} \varphi^{\prime} \, {\rm Im}(\ee^{-i \alpha} { W})
\right\}\,,
\end{align}
where we discarded a total derivative. We note here that ${\cal E}$, originally defined in \eqref{E-def}, now contains the multiplier $\ee^\varphi$ due to the rescaling in \eqref{static-rescal} above.

Note that since the addition of the Lagrange multiplier $\ee^\varphi$ in the original action leads
to an ungauged theory, it is possible at this stage to proceed in solving the equations of
motion by simply putting $\ee^\varphi=0$, which is a consistent solution that eliminates all instances of the vector of gaugings. However, it is clear that it is not necessary to make this choice for this function a priori.
Indeed, making instead a choice for the metric function $\ee^\psi=r$, so that the base metric in \eqref{4dmetric} is flat three dimensional space, one can obtain a more general squaring of the ungauged bosonic action. In this case, the kinetic term for the function $\ee^\psi$ is trivial and the action can be further rearranged into
\begin{align}
S_{1d}
=\int dr &\left\{ \, -\frac12 \ee^{2(U-\psi)} \Iprod{\cal E}{\mathrm{J}\,{\cal E}}
 -\ee^{2\psi}\left[ (Q_r + \alpha')
 + 2 \ee^{-U} \, {\rm Re}(\ee^{-i \alpha} { W}) \right]^{2}\right.
\nonumber\\
& \quad
 -\left[2\,r\,\ee^{\varphi-U}\, {\rm Im}(\ee^{-i \alpha} W)
     -(1 +\tfrac12\, r\, \varphi^{\prime}) \right]^{2}
\nonumber\\
& \quad \left.
 +r^{4}\ee^{\varphi} \left( (r^{-1}\ee^{-\varphi/2})^{\prime}\right)^2
 -\left(2+\ee^{\varphi}\,\langle G,\Gamma \rangle\right)
   \right\}\,,
\end{align}
which is manifestly a sum of squares for the physical fields, along with an extra
kinetic term and a Liouville-type potential for the multiplier $\varphi$, that
decouples from the rest of the action.

One can now solve the equations of motion for the physical
fields by imposing that each of the squares vanishes, as
\begin{eqnarray}
{\cal E} &=& 0\,, \label{eq:scal-flow}\\
Q_r + \alpha'&=& -2\ee^{\varphi} \ee^{-U}\,\,{\rm Re}(\ee^{-i \alpha} W)\,, \label{eq:Kah-1}\\
2\,r\,\ee^{-U}\, {\rm Im}(\ee^{-i \alpha} W)
    &=& \ee^{-\varphi}(1 +\tfrac12\, r\, \varphi^{\prime})\,. \label{eq:phiprime}
\end{eqnarray}
These equations describe the flow of the scalars and the scale factor $\ee^U$, as well
as fix the function $\ee^\varphi$ in terms of physical fields. In addition, one still has
to impose the Hamiltonian constraint \eqref{eq:Ham-constraint} above, as well as the  equation
of motion for the Lagrange multiplier $\varphi$, which reads
\begin{equation}\label{eq:Lagrange-eom}
 \frac{d}{dr}\left( r^2\, u^{\prime} \right) - \Iprod{\cG}{\Gamma}\,r^{-2}\,\ee^{-2u}=0\,,
\end{equation}
where we used the variable $\ee^u\equiv r^{-1}\ee^{-\varphi/2}$ for convenience.

The flow equations above are closely related to the ones obtained in \cite{Dall'Agata:2010gj} for
gauged supergravity, with the difference that the function $\ee^\psi$ describing the spatial
part of the metric is now fixed to $\ee^\psi=r$ and that we have included the additional
function $\ee^\varphi$. One can decompose the scalar flow equations \eqref{eq:scal-flow}
in components to find
\begin{align}
\label{eq:metr-factor}
U' =&\, - r^{-2}\,\ee^{U} \, {\rm Re}(\ee^{-i \alpha}Z)
     +\ee^\varphi \ee^{-U}\,{\rm Im}(\ee^{-i \alpha} W)\,,
\\
\label{eq:phys-scalars}
t^i{}' =&\, - \ee^{i \alpha} g^{i \bar \jmath}\left(\ee^{U-2 \psi}\bar{Z}_{\bar \jmath}
   + i\,\ee^\varphi \ee^{-U} \bar{W}_{\bar \jmath}\right)\,,
\end{align}
along with one more equation for the K\"ahler connection
\begin{equation} \label{eq:Kah-2}
Q_r + \alpha' = - r^{-2}\,\ee^{U}\, {\rm Im}(\ee^{-i \alpha}Z)
       - \ee^{\varphi} \ee^{-U}\,\,{\rm Re}(\ee^{-i \alpha} W) \,.
\end{equation}
Combining the last relation with \eqref{eq:Kah-1} leads to the constraint
\begin{equation}\label{eq:Z-L-constraint}
 r^{-2}\,\ee^{U}\, {\rm Im}(\ee^{-i \alpha}Z)
       = \ee^{\varphi} \ee^{-U}\,\,{\rm Re}(\ee^{-i \alpha} W)\,,
\end{equation}
which in the case of genuinely gauged supergravity in \cite{Dall'Agata:2010gj},
can be shown to be equivalent to the Hamiltonian constraint \eqref{eq:Ham-constraint}.
However, for the theory at hand, \eqref{eq:Ham-constraint} is not automatically
satisfied upon using \eqref{eq:metr-factor}-\eqref{eq:Z-L-constraint}, but takes the form
\begin{equation}\label{eq:Ham-const-fin}
\Iprod{G}{\Gamma}
+  4 r^{2}\,\ee^{-2U} \ee^{\varphi} \, \left({\rm Im} (\ee^{- i \alpha}  W )\right)^2=0 \,,
\end{equation}
which relates $\ee^\varphi$ to the physical fields.

\subsection{Asymptotically flat solutions}
\label{asym-flat-stat}

We can now look for solutions to the above system, starting with the observation that \eqref{eq:Lagrange-eom} can be solved explicitly. The general solution can be written in terms of exponentials of the type $\ee^{\pm 1/r}$, which are badly singular at $r=0$ and lead to unphysical results. However, this differential equation also has the particular enveloping solution
\begin{equation}\label{eq:Lagrange-sol}
\ee^{u}\equiv r^{-1}\ee^{-\varphi/2}
=\Iprod{\Gamma}{\cG}^{-1/2}\,\left(\mathrm{v}+\tfrac{\Iprod{\Gamma}{\cG}}r\right)
=\Iprod{\Gamma}{\cG}^{-1/2}\, V\,,
\end{equation}
where $\mathrm{v}$ is a constant and we assumed that $\Iprod{\Gamma}{\cG}\!>\!0$, so that the distinguished harmonic function $V$ defined above is positive definite.
From \eqref{eq:Ham-const-fin}, we obtain
\begin{equation}\label{eq:ImL}
 2 \ee^{-U} \,{\rm Im} (\ee^{- i \alpha}  W )=V\,,
\end{equation}
where the positive root was chosen by imposing \eqref{eq:phiprime}. The last relation implies that the solution can be expressed in terms of harmonic functions, as shown in \cite{Bossard:2012xs}.

Indeed, the flow equations \eqref{eq:metr-factor} and \eqref{eq:phys-scalars} with the particular solution for $\varphi$ given by \eqref{eq:Lagrange-sol}, can be straightforwardly shown to be identical to the static limit of the flow equations derived in \cite{Bossard:2012xs} for the single centre class of asymptotically flat black holes, upon identifying the gaugings $\cG$ with the auxiliary vector $\hat{R}^*$ used to express the solution\footnote{The interested reader can find an outline of this identification in appendix \ref{app:rot}, where the full rotating single center class is considered.}.
Note that in \cite{Bossard:2012xs} the auxiliary vector $\hat{R}^*$ was required to be very small by consistency of the Einstein equations for asymptotically flat black holes. Moreover, in that work it was found that regularity requires an additional constraint on the system, which can be expressed in several equivalent ways. In terms of the scalars, this constraint takes the form of the reality condition
\begin{equation} \label{RealT}
e^{-i  \alpha} dt^i - i\,\ee^{\cal K}c^{ijk}\, \hat W_j\, g_{k \bar k} d \bar t^{\, \bar k}
+ \hat W\, \overline{\hat{W}}{}^i \, \overline{\hat{W}}{}_{\bar \imath} d \bar t^{\, \bar \imath}
= \overline{\hat{W}}{}^i \left(1 + e^{-i  \alpha}  \hat W   \right)\,  dU\,,
\end{equation}
where we defined the following shorthand expressions for convenience
\begin{equation}
 \hat W= |W|^{-1}\, W\,, \qquad  \hat W_a = |W|^{-1} \, e_a^i W_i\,.
\end{equation}
The reality condition \eqref{RealT} can be used to show the existence of a second constant very small vector throughout the flow, given by
\begin{align}
R = -4\,\frac{\ee^{-2U}}{|Y|^2\,V^2}&\,
\mbox{Re}\Bigl[
 Y^3\,\bar W\,{\mathcal V}
+ |Y|^2 \,Y\,\bar W^{i}\, D_i \cV \Bigr]\,,
\label{Rstar} \\
Y \equiv &\,  ( 1 + i\,\mathrm{m}\,e^{2U}) \label{Y-def}\,.
\end{align}
Here, $\mathrm{m}$ is an arbitrary constant that is promoted to a dipole harmonic function in the rotating case (see appendix \ref{app:rot}).
This vector can be shown to be mutually local with $\Gamma$, $\Iprod{R}{\Gamma}=0$, using the flow equations above, but is nonlocal with $\cG$, as $\Iprod{\cG}{R}=-4$, and in simple cases it can be viewed as the magnetic dual of $\cG$. Alternatively, one can derive the constraint \eqref{RealT} by demanding that the vector $R$ be constant.

Given the definitions above, the solution to the system \eqref{eq:metr-factor}-\eqref{eq:Kah-2} is given by
\begin{equation}
2\,\I(e^{-U} e^{-i \alpha} \cV)=
 \cH - 2\,\tfrac{\Iprod{\cG}{\cH}}{\Iprod{\cG}{R}}\,R
 + \tfrac{\mathrm{m}}{\Iprod{\cG}{\cH}}\, \cG \,,
\label{scal-sol}
\end{equation}
where $\cH$ are harmonic functions carrying the charges as
\begin{equation}
 \cH=\mathrm{h} + \frac{\Gamma}{r}\,,
\end{equation}
and the distinguished harmonic function $\Iprod{\cG}{\cH}$ is fixed by \eqref{eq:ImL} as
\begin{equation}
\Iprod{\cG}{\cH}= -V\,.
\end{equation}

The reality constraint \eqref{RealT} can now be recast in terms of the harmonic functions describing the solution. Using the flow equations \eqref{eq:metr-factor} and \eqref{eq:phys-scalars} to express the derivatives of the scalars in terms of the gauge fields, one can show that an equivalent form of the same reality condition can be obtained, as
\begin{equation} \label{sympl-constr}
\frac12\, I^{'\,M}_{4}(\cH,\cG) = \Iprod{\cG}{\cH}\, \cH^M
- 2 \frac{\Iprod{\cG}{\cH}^2 }{\Iprod{\cG}{R}} R^M \, .
\end{equation}
Here, we use the index $M,N,\dots$ to denote both electric and magnetic components and $I^{'\,M}_{4}$ is the derivative of the quartic invariant, defined in terms of a completely symmetric tensor $t^{MNPQ}$ as
\begin{align}  \label{I4-der}
I_4(\cH)\equiv & \frac{1}{4!} t^{MNPQ}\cH_M\cH_N\cH_P\cH_Q
\nonumber\\
 I^{'\,M}_{4}(\cH,\cG) \equiv & \frac{\partial^2 I_4(\cH)}{\partial \cH_M \partial\cH_N} G_N
= \frac{1}{2} t^{MNPQ} \cH_N \cH_P G_Q \,.
\end{align}
In \cite{Bossard:2012xs} it was shown that if $\cG$, and thus $R$, are very small, this constraint implies that the harmonic functions $\cH$ lie in a Lagrangian submanifold that includes $R$. Near the horizon, one finds the same constraint for the charges, so that a particular choice of $G$ restricts the physical charges to lie in the same Lagrangian submanifold. We refer the interested reader to that work for details on the derivation of these results.

This concludes our analysis of the embedding of extremal asymptotically flat
black holes in gauged supergravity for the static case. We refer to appendix
\ref{app:rot} for a similar analysis in the rotating case. It turns out that
the inclusion of a Lagrange multiplier in exactly the same way leads
to the same equation of motion \eqref{eq:Lagrange-eom} and the same solution
\eqref{eq:Lagrange-sol} as above. The result is an extension of the static
embedding of this section to the most general asymptotically flat extremal
under-rotating black holes, as obtained in \cite{Bossard:2012xs}. The solution turns out to take exactly the same form as in \eqref{scal-sol} with the constant $\mathrm{m}$ replaced by a dipole harmonic function describing the rotation.

\section{BPS attractors in abelian gauged theories}
\label{sec:BPShor}

As already announced in the introduction, the embedding of asymptotically flat black holes in the flat gauged theories we consider in this paper allows to show that their near-horizon geometries are in fact supersymmetric. In the previous section we saw a close similarity between the static flow equations for asymptotically flat and $1/4$-BPS black holes in AdS. Below we further establish that static horizons in both Minkowski and AdS spaces in fact belong to a common $1/2$ BPS class of solutions\footnote{This is in accordance with our results in section 2. From this point of view, the crucial factor that allows for a
unified discussion is that the expression for the Lagrange multiplier reduces to a constant in
the near horizon region (cf. the solution in \eqref{eq:Lagrange-sol}), thus diminishing any difference between the flat and AdS case. This is the case even for under-rotating black holes, as shown in appendix \ref{app:rot}.} already discussed in \cite{deWit:2011gk}. Beyond the static class, we further analyze the near-horizon geometry of extremal under-rotating black holes \cite{Astefanesei:2006dd}, whose flow equations are discussed in appendix \ref{app:rot}. These turn out to preserve $1/4$ of the supercharges, which completes the statement that all asymptotically flat static and under-rotating extremal black holes have BPS horizons.

In order to study supersymmetric solutions, we only need to explicitly ensure that the supersymmetry variations of the fermions vanish. All supersymmetry variations
for the bosons are automatically zero by the assumption of vanishing fermions.
The fermionic fields that belong to the supermultiplets appearing in the action \eqref{Ssugra4D}
are the gravitini $\psi_{\mu A}$ for the gravity multiplet and the gaugini $\lambda^{iA}$ for the
vector multiplets. The corresponding supersymmetry variations are\footnote{Here we choose to orient the FI terms along direction 3 of the
quaternionic moment maps, as done in \cite{Hristov:2010ri}.}:
 \begin{align}\label{susy-vars}
\nonumber \delta \psi_{\mu A}&=
 D_\mu \varepsilon_A
- 2 i\,X^I\, I_{I J}\, F^{J-}_{\mu\nu}\,\gamma^\nu\, \epsilon_{AB}\varepsilon^B
- \tfrac 12\, W\, \sigma^3_{A B} \gamma_\mu\,\varepsilon^B\,,
\\
  \delta \lambda^{iA}&= -i\slash{\partial} t^i \,\varepsilon^A
 - \bar{D}^i\bar X^I\,
  I_{I J}\, F^{J-}_{\mu\nu}\,\gamma^{\mu\nu}\,\epsilon^{AB}\varepsilon_B
+ i\,\bar W^{i} \, \sigma^{3, AB} \varepsilon_B\,,
\end{align}
where the covariant derivative $D_\mu$ reads
\begin{align}\label{cov_der}
D_\mu \varepsilon_A = \big(\partial_\mu - \tfrac 14 \omega_\mu^{ab}
  \gamma_{ab} + \tfrac i 2 Q_\mu \big)\varepsilon_A +
  \tfrac{i}2\,\Iprod{G}{A_\mu}\, \sigma^3_{A}{}^B \varepsilon_B\,,
\end{align}
and $W$ and $W_{i}$ are the central charges defined in \eqref{eq:W-def}. The symplectic product $\Iprod{G}{A_\mu}$ in the standard electrically gauged supergravity just involves the electric gauge fields $A^{\Lambda}_{\mu}$ \cite{Andrianopoli:1996cm}, but has a straightforward generalization, as shown in \cite{deWit:2005ub,deWit:2011gk}.
We also used the shorthand $I_{IJ}=\mbox{Im}\cN_{IJ}$ for the period matrix. The presence of this
matrix seems to spoil duality covariance on first sight, but it is possible to rewrite
the relevant terms in a form convenient for our purposes, as in
\cite{Ceresole:1995ca}
 \begin{align}\label{susy-vars-new}
\nonumber \delta \psi_{\mu A}&=
 D_\mu \varepsilon_A
+ Z(\cF)_{\mu\nu}^{-}\,\gamma^\nu\, \epsilon_{AB}\varepsilon^B
- \tfrac 12\, W\, \sigma^3_{A B} \gamma_\mu\,\varepsilon^B\,,
\\
  \delta \lambda^{iA}&= -i\slash{\partial} t^i \,\varepsilon^A
 + \tfrac{i}2\,\bar Z(\cF)^{i\,-}_{\mu\nu}\,\gamma^{\mu\nu}\,\epsilon^{AB}\varepsilon_B
+ i\,\bar W^{i} \, \sigma^{3, AB} \varepsilon_B\,.
\end{align}
Here, the central charges of the electric and magnetic field strengths are
computed component-wise as in \eqref{ch-def}, as
\begin{equation}
 Z(\cF)_{\mu\nu}=\Iprod{\cF_{\mu\nu}}{\cV}\,, \qquad
Z(\cF)_{i, \mu\nu}=\Iprod{\cF_{\mu\nu}}{D_i\cV}\,,
\end{equation}
which are already anti-selfdual and selfdual respectively due to \eqref{cmplx-sdual} and
\eqref{CY-hodge}.

Note that the FI parameters $G$ above are assumed to be generic, and include the particular
choice of FI parameters such that the scalar potential is identically zero. In that limit, we obtain a theory with a
bosonic Lagrangian identical to ungauged supergravity, but with a different fermionic sector
that involves a nonzero very small vector of FI terms explicitly. Therefore, the supersymmetry variations above are
strictly valid only for gauged supergravity, even though the associated bosonic backgrounds
we describe below are solutions to both gauged and ungauged supergravity. The supersymmetric solutions of the two theories however do not overlap and form two disjoint sets. This is easy to see because the ungauged supersymmetry variations are again given by \eqref{susy-vars-new} after setting $G = 0$, leading to the vanishing of $W, W^i$. Suppose now that we have a supersymmetric background solution of the ungauged theory and let us focus for simplicity on the gaugino variation. If we also want it to be a solution of the flat gauged theory, we require that it automatically satisfies $W^i = 0$, since otherwise one cannot make the variation vanish both in the gauged and in the ungauged theory. Now, using the vanishing of the scalar potential \eqref{gau-pot} we find
$$ |W|^2 = \tfrac13\,W_i \bar{W}^i \ ,$$
which means that we also need $W = 0$ for the hypothetical BPS solution in both theories. However, it is a special geometry property that
\begin{equation}
\begin{pmatrix} \Iprod{\cV}{G}\\ \Iprod{D_i\cV}{G}\end{pmatrix} = 0 \,, \quad \Rightarrow\quad G=0\,,
\end{equation}
since one can invert the matrix multiplying $G$ in this equation. This leads us back to the ungauged case, and we find a contradiction. Therefore every BPS solution of the ungauged theory (e.g.\ the asymptotic Minkowski spacetime connected to the asymptotically flat black holes) is not supersymmetric in the flat gauged theory, and vice versa (e.g.\ the black hole attractor geometries are BPS in the gauged theory, as shown below, but break supersymmetry in the ungauged theory) as schematically illustrated by Fig. 1.

We now move on to the explicit analysis of the supersymmetries preserved by the various horizon geometries. In doing so, we will be using a timelike Killing spinor ansatz, ensuring that once the BPS equations hold we already have supersymmetric solutions, i.e.\ the BPS equations together with the Maxwell equations and Bianchi identities imply the validity of the Einstein and scalar equations of motion (see \cite{Kallosh:1993wx,Hristov:2010eu}). This is important for the discussion of backgrounds with non-constant scalars, which are the ones relevant for rotating attractors.

In section \ref{sec-stat-attr} we verify that the attractor equations obtained as a limit of the full $1/4$-BPS static solutions in AdS$_4$ in \cite{Dall'Agata:2010gj}, do exhibit supersymmetry enhancement to $4$ real supercharges. We then identify the attractor equations of static asymptotically flat non-BPS black holes of \cite{Bossard:2012xs} as a subset of the BPS attractors in gauged supergravity, in the limit of flat gauging where the FI terms are restricted to be a very small vector. Similarly, in section \ref{sec-rot-attr} we show that the general under-rotating attractor solutions of \cite{Bossard:2012xs} preserve $1/4$ of the supersymmetries.

\subsection{Static attractors}
\label{sec-stat-attr}

We first concentrate on the near horizon solutions of static black holes, therefore we consider metrics of the direct product form AdS$_2 \times$S$^2$ with radii $v_1$ and $v_2$ of AdS$_2$ and S$^2$, respectively:
\begin{equation}\label{6:metric-ansatz}
{\rm d} s^2 = -\frac{r^2}{v_1^2}\, {\rm d}t^2 + \frac{v_1^2}{r^2}\, {\rm d}r^2
 + v_2^2\, ({\rm d} \theta^2 + \sin^2 \theta {\rm d} \phi^2)\ .
\end{equation}
The corresponding vielbein reads
\begin{equation}
e_{\mu}^a = {\rm diag}\Big(\frac{r}{v_1}, \frac{v_1}{r}, v_2, v_2 \sin
\theta\Big)\ ,
\end{equation}
whereas the non-vanishing components of the spin connection turn out to
be
\begin{equation}
\omega_t^{0 1} = -\frac{r}{v_1^2}, \qquad \omega_{\phi}^{23} = \cos \theta\ .
\end{equation}
We further assume that the gauge field strengths are given in terms of the charges $\Gamma=(p^I\,, q_I)^T$ by
\begin{equation} \label{6:el_field_strengths}
\mathcal{F_{\mu \nu}} \equiv (F^I_{\mu \nu}; G_{I}{}_{\mu \nu})\ , \qquad
F^{I}_{\theta \varphi}
=\tfrac{1}{2}\,p^{I} \,\sin \theta\,,
\qquad
G_{I}{}_{\theta \phi}
=\tfrac{1}{2}\,q_{I}\, \sin \theta\,,
\end{equation}
which are needed in the BPS equations below. The scalars are assumed to be constant everywhere, $\partial_{\mu} z = 0$, as always on the horizon of static black holes. This ansatz for gauge fields and scalars automatically solves the Maxwell equations and Bianchi identities in full analogy to the case of ungauged supergravity.

Anticipating that the near horizon geometries of the solutions described in the previous
section preserve half of the supersymmetries, we need to impose a projection
on the Killing spinor. This is in accordance with the fact that these solutions cannot be
fully supersymmetric once we require that not all FI terms vanish (see
\cite{Hristov:2009uj, Louis:2012ux} for all fully BPS solutions in $\cN\!=\!2$ theories in $4d$).
Taking into account spherical symmetry, there are only two possibilities in
an AdS$_2 \times$S$^2$ attractor geometry, as shown in \cite{deWit:2011gk}. Namely, one either
has full supersymmetry, and therefore no projection is involved, or $1/2$-BPS geometries
satisfying the projection
\begin{equation}\label{6:KS-ansatz}
\varepsilon_A = i {\sigma}^3_A{}^B\,\, {\gamma}^{23}\, \varepsilon_B 
 =  {\sigma}^3_A{}^B\,\, {\gamma}^{01}\, \varepsilon_B\ ,
\end{equation}
with the last equality due to the fact that spinors are chiral in the chosen conventions
(these are exhaustively listed in \cite{Andrianopoli:1996cm,Hristov:2012bk}). Note that
a Killing spinor satisfying this projection is rather different from the standard
timelike Killing spinor projection that appears in asymptotically flat 1/2-BPS solutions
(shown in \eqref{KS-time} below, see e.g.~\cite{Behrndt:1997ny}), but is exactly the same
as one of the projections appearing in asymptotically AdS$_4$ 1/4-BPS solutions (see \cite{Dall'Agata:2010gj,Hristov:2010ri}).

\subsubsection*{Analysis of the BPS conditions}

Now we have all the data needed to explicitly write down the supersymmetry variations of the gravitini and gaugini. To a certain extent this analysis was carried out in section 8 of \cite{deWit:2011gk} and will not be exhaustively repeated here. One can essentially think of the Killing spinors as separating in two - a part on AdS$_2$ and another part on S$^2$. It turns out that the AdS$_2$ part transforms in the standard way under the $SO(2,1)$ isometries of the AdS space, while the spherical part remains a scalar under rotations. The $t$ and $r$ components of the gravitino variation are therefore non-trivial due to the dependence of the spinor on these coordinates. We are however not directly interested in the explicit dependence, but only consider the integrability condition for a solution to exist, given by $D\!_{[t}D\!_{r]} \varepsilon_A=0$ for all $A = 1,2$.
Plugging the metric and gauge field ansatz, this results in the equations
\begin{equation}\label{cond1}
\frac{1}{2 v_1^2}
 = |W|^2+\frac{1}{v_2^4}\,|Z|^2\,,
\qquad
\Iprod{G}{\cF_{tr}}=0\,.
\end{equation}
The solution of this equation therefore ensures the vanishing of the gravitino variation on AdS$_2$. Turning to the spherical part, with the choice of Killing spinor ansatz it is easy to derive two independent equations that already follow trivially from the analysis of \cite{Hristov:2010ri},
\begin{equation}\label{cond2}
i\,\frac{1}{v_2^2}\,Z = - W\ ,
\end{equation}
and
\begin{equation}\label{Dirac_quant}
\Iprod{G}{\Gamma} = -1\ ,
\end{equation}
which is the usual Dirac quantization condition\footnote{From the point of view of the flow equations derived in section 2, $\Iprod{\Gamma}{\cG}$ can be an arbitrary non-vanishing constant. This is exactly the value of the Lagrange multiplier in \eqref{eq:Lagrange-sol} at the horizon, thus rescaling the gauging vector as $\cG^\prime=\Iprod{\Gamma}{\cG}^{-1}\cG$ in the solution \eqref{scal-sol} in that limit. It follows that $\Iprod{\cG^\prime}{\Gamma}=-1$, which is the choice made in \cite{Cacciatori:2009iz,Dall'Agata:2010gj, Hristov:2010ri} for the full solution and we adopt
it here, dropping the primes on the gaugings, to make our notation in sections 2 and 3 consistent without any loss of generality.} that seems to accompany the solutions of "magnetic" type\footnote{The solution at hand, called magnetic AdS$_2 \times$S$^2$ in \cite{Hristov:2012bk}, is the near horizon geometry of asymptotically magnetic AdS$_4$ black holes \cite{Hristov:2011ye,Hristov:2011qr}.}. Note that \eqref{cond2} can be used to simplify the first of \eqref{cond1}, so that we can cast the above conditions in a more suggestive form for our purposes, as
\begin{align}\label{cond4}
\begin{split}
v_1^{-2} 
= 4\, |W|^2 \ ,\qquad
v_2^2 = -i\, \frac{Z}{W}\ .
\end{split}
\end{align}

Moving on to the gaugino variation, the condition that the scalars remain constant leaves us with only one (for each scalar) additional condition on the background solution,
\begin{equation}\label{cond3}
-i\,Z_i = v_2^2\, W_i  \ .
\end{equation}
This concludes the general part of our analysis - it turns out that in FI gauged supergravity one can ensure that AdS$_2 \times$S$^2$ with radii $v_1$ and $v_2$ preserves half of the supersymmetries by satisfying equations \eqref{Dirac_quant}-\eqref{cond3} within the metric and gauge field ansatz chosen above. These equations are in agreement with the analysis of \cite{Bellucci:2008cb, Dall'Agata:2010gj}. Moreover, the attractors above are a realisation of the 1/2-BPS class of AdS$_2\times$S$^2$ vacua of \cite{deWit:2011gk}, which are described by the superalgebra $SU(1,1|1)\times SO(3)$, as opposed to the fully BPS AdS$_2\times$S$^2$ vacua that are described by $SU(1,1|2)$.

The above equations can be written in terms of symplectic vectors (e.g.\ as in \cite{Dall'Agata:2010gj}), so that they can be directly compared to \cite{Bossard:2012xs}. To this end, one can straightforwardly see that the condition
\begin{equation}\label{univ-attr}
-4\,{\mathrm{Im}}(\bar Z {\cal V})
= \Gamma + v_2^{2}\,\mathrm{J} \cG \,,
\end{equation}
is equivalent to \eqref{cond2} and \eqref{cond3}, while the first of \eqref{cond4}
has to be used to fix the AdS$_2$ radius. Alternatively, one may write the attractor
equations by solving for ${\cal V}$ in terms of the charges and gaugings. Since all
the above equations are invariant under K\"ahler transformations, we need to introduce
an a priori arbitrary local phase $\ee^{i\alpha}$, which is defined to have unit K\"ahler
weight. One can then combine \eqref{cond1} and \eqref{cond3} to obtain
\begin{equation}\label{univ-attr-mod}
2\,\frac{v^2_2}{v_1}\,{\mathrm{Im}}(\ee^{-i\alpha} {\cal V})
= \Gamma + v_2^{2}\,\mathrm{J} \cG \,,
\end{equation}
while \eqref{cond2} has to be viewed as an additional constraint. Taking the inner
product of \eqref{univ-attr-mod} with $\Gamma + v_2^{2}\,\mathrm{J} \cG$ identifies
the phase $\ee^{i\alpha}$ as the phase of the combination in \eqref{cond2}, which
drops out from that relation.

In order to show that these BPS conditions above do indeed admit solutions describing
asymptotically flat black holes, one can consider
the inner product of \eqref{univ-attr} with the gaugings, using \eqref{Dirac_quant},
to show that the sphere radius is given by
\begin{align}
 v_2^{-2} = 2\,g^{i \bar \jmath}\,{W}_{i}\,\overline{W}_{\bar \jmath} -2\,| W|^2\,.
\end{align}
Upon imposing triviality of the potential as in \eqref{eq:doub-crit}, the above
expression and the first of \eqref{cond4} imply that $v_2\!=\!v_1$, which is necessary
for asymptotically flat black holes. Indeed, using the definition \eqref{Rstar}
in this special case, the generic BPS attractor equation in the form \eqref{univ-attr-mod}
can be written as
\begin{equation}
2\,v_1\,{\mathrm{Im}}(\ee^{-i\alpha} {\cal V})
=\Gamma + \tfrac12\, R \,.
\end{equation}
These are exactly the general attractor equations for asymptotically flat
black holes found in \cite{Bossard:2012xs} for the ungauged
case\footnote{We remind the reader that the inner product
$\Iprod{\Gamma}{G}$ has to be rescaled to unity in the original reference for a proper comparison with this section.}.
We conclude that the near horizon region of static asymptotically flat extremal black holes can be viewed as a special
case of the general attractor geometry for BPS black holes in abelian gauged supergravity, upon restricting the FI  parameters to be a very small vector, thus leading to a flat potential.

In addition, when all FI parameters are set to zero, one immediately obtains the BPS attractor equations
of ungauged supergravity, preserving full $\cN\!=\!2$ supersymmetry
\cite{Ferrara:1995ih, Strominger:1996kf, Ferrara:1996dd}. This provides us with a unifying
picture, since the BPS attractor equations \eqref{univ-attr} appear to be
universal for static extremal black holes in $\cN\!=\!2$ theories, independent of the asymptotic behavior (Minkowski or AdS) or the amount of supersymmetry preserved.

One intriguing aspect of this result is that, while in the ungauged theory ($G=0$), the attractor equation leads to a well defined metric only when the quartic invariant of the charges, $I_4(\Gamma)$, is positive, the presence of a nontrivial $G$ does not seem to allow for a charge vector $\Gamma$ with a positive quartic invariant, i.e.~in all known examples $I_4<0$ iff $G\neq 0$, both for asymptotically flat and AdS black holes. Similarly, the explicit AdS$_4$ solutions of \cite{Cacciatori:2009iz, Dall'Agata:2010gj, Hristov:2010ri}, also have a negative quartic invariant of the charges, contrary to the intuition one might have from the asymptotically flat case. It is natural to expect that the quartic invariant of charges allowed for asymptotically AdS$_4$ BPS solutions is negative even though this is not the only quantity that controls the horizon in that case.

In view of the above, it is interesting at this point to make some comments on the potential microscopic counting of degrees of freedom, which can be now safely discussed due to the presence of supercharges on the horizon. From a microscopic string theory perspective we know that the FI parameters are usually some particular constants corresponding to topological invariants of the compactification manifolds, see \cite{Cassani:2012pj} for a clear overview and further references. This means that one is not free to tune the value of the vector $G$. We further know that one of the electromagnetic charges is uniquely fixed by the choice of $G$, meaning that we are not free to take the large charge limit in this particular case. We then find that the black hole entropy, $\mathrm{S}$, which is proportional to the area of the horizon, scales as $\mathrm{S}\sim\Gamma^{3/2}$, a behavior that is in between the usual $\mathrm{S}\sim\Gamma^2$ of 1/2 BPS asymptotically flat black holes\footnote{Note however, that the
entropy of asymptotically flat $1/2$-BPS black holes in five dimensions scales exactly as $\Gamma^{3/2}$, see e.g.\ \cite{Strominger:1996sh}.} and the $\mathrm{S}\sim\Gamma$ case of 1/4 BPS asymptotically magnetic AdS black holes \cite{Cacciatori:2009iz,Hristov:2010ri}. This is of course not a puzzle on the supergravity side, where we know that some charges are restricted, but it provides a nontrivial check on any potential microscopic descriptions of black hole states in string theory.

\subsection{Under-rotating attractors}
\label{sec-rot-attr}

We now turn to the more general case of extremal under-rotating attractors
corresponding to asymptotically flat solutions \cite{Astefanesei:2006dd}. These are
described by a more general fibration of S$^2$ over AdS$_2$ that incorporates rotation as
\begin{align}\label{rot-attr-metr}
  ds^2 = -e^{2U}\,r^2\,(d t+\omega)^2
 + e^{-2U} &\,\left( \frac{dr^2}{r^2} + d\theta^2 + \sin^2{\theta} d\phi^2 \right)\,,
\nonumber\\
  \ee^{-4U}= -I_4(\Gamma) - \mathrm{j}^2\cos^2\theta &\,,
\qquad
\omega= \mathrm{j}\,\frac{\sin^2\theta}{r}\,d \phi\,,
\end{align}
where $\mathrm{j}$ is the asymptotic angular momentum. It is easy to see that this metric
reduces to \eqref{6:metric-ansatz} for $v^2_1=v^2_2=\sqrt{-I_4(\Gamma)}$ upon setting $\mathrm{j}=0$ above. We choose the vielbein
\begin{equation}
e_t^0= r e^U\,,\quad e_r^1 = \frac{e^{-U}}{r}\,,\quad e_{\theta}^2 = e^{-U}\,,
\quad e_{\phi}^3 = e^{-U} \sin \theta\,,\quad e_{\phi}^0 = j\, e^U \sin^2 \theta\ ,
\end{equation}
which leads to the following non-vanishing components of the spin connection:
\begin{eqnarray}
&\omega^{0 1} = -r\, e^{2 U} (dt + \tfrac12\,\omega ), \quad
&\omega^{2 3} = -e^{4 U} \cos \theta\, \left(j\, r\, dt + \tilde{v}\,d\phi \right) \ ,
\\
&\omega^{0 2} = j\, r\, e^{6 U} \sin \theta \cos \theta\, (\tfrac{j}{2}r dt + \tilde{v} d\phi)\,, \quad
&\omega^{0 3} = -\tfrac{j}{2r}\, e^{2 U} \,( \sin\theta\,dt - r\,\cos{\theta}d\phi)\,,
\\
&\omega^{1 2} = -\frac{1}{2 r}\,j^2\, e^{4 U} \sin \theta \cos \theta\, dr\,, \quad
&\omega^{1 3} = \tfrac{1}{2}\,j\, r\, e^{4 U} \sin \theta (dt + \omega )\,,
\end{eqnarray}
where we defined the function
\begin{equation}
 \tilde{v} = I_4 - \tfrac12\,j^2 (1 + \cos^2 \theta)\ .
\end{equation}
The gauge fields for this class of solutions read \cite{Bossard:2012xs}
\begin{align}\label{gauge_fields_rot}
 \cF= d\big[\zeta\,r\,(dt + \omega)\big] +&\, \Gamma\, \sin\theta\,d\theta\wedge d\phi\,,
\qquad
\zeta = -2\,e^{U} \R [ e^{-i \alpha } \cV ] + G\,,
\end{align}
where we used the fact that the section depends on the radial coordinate by
an overall $r^{-1}$ in the near horizon region, as for the static case.
We refrain from giving the full solution for the scalars at this stage, since it will
be derived from the BPS conditions below. Here we note that the physical
scalars $t^i$ only depend on the angular coordinate $\theta$ in the near horizon
region, and we give the expression for the K\"ahler connection
\begin{equation}
 Q+d \alpha= \tfrac12\,\mathrm{j}\,\ee^{2U}\,\sin\theta\,d\theta\,,
\end{equation}
for later reference. The interested reader can find an explicit example solution t the STU model in
Appendix \ref{app:example}, both at the attractor and for the full flow.

As already mentioned above, the backgrounds we are interested in only preserve
two supercharges, i.e. they are $1/4$-BPS. The fact that we now need a second projection on the Killing spinor, in addition to \eqref{6:KS-ansatz}, can be derived directly by considering the BPS equations, e.g.\ the gaugino variation. We omit details of this derivation, which is straightforward, and just give the resulting additional projection
\begin{equation}\label{KS-time}
{\gamma}^{0}\,\varepsilon_A = i\, \ee^{i\alpha} \epsilon_{AB}\,\varepsilon^B \ ,
\end{equation}
which is the same as the one used in e.g.~\cite{Behrndt:1997ny, Hristov:2010eu, Dall'Agata:2010gj, Hristov:2010ri}.

As in the static case, we make use of the complex self-duality of $\cF$, so
that we only need to use half of its components. We therefore choose for
convenience $\cF_{0\hat a}$ and $\cF_{23}$, where $\hat a=2,3$ is a
flat index on the sphere, given by
\begin{align}\label{F-comp-rot}
 \cF_{23}=&\, \ee^{2U}\,\Gamma + \zeta\,r\,(d \omega)_{23}
     = \ee^{2U}\,\left(\Gamma + 2\,\zeta\,\mathrm{j}\,\cos\theta\right) \,,
\qquad
 \cF_{0\hat a}= - \partial_{\hat a}\zeta\,.
\end{align}
Since the central charges of these quantities appear in the BPS conditions,
we note for clarity the following relations
\begin{align}
Z(\zeta)= -i\,\ee^U\ee^{i\alpha} + W\,,
\qquad &\,
 Z_i(\zeta)= W_i\,,
\\
 Z(\partial_{\hat a}\zeta)=
-i\,\ee^U\ee^{i\alpha}\big[ \partial_{\hat a}U + i(Q_{\hat a}+\partial_{\hat a}\alpha) \big]\,,
&\,\qquad
Z_i(\partial_{\hat a}\zeta)= i\,\ee^U\ee^{i\alpha}g_{i {\bar \jmath}} \partial_{\hat a}\bar t^{\bar \jmath}\, \,,
\label{Z-zeta}
\end{align}
which can be straightforwardly derived from \eqref{gauge_fields_rot}
using \eqref{eq:D-gauge}.

\subsubsection*{Analysis of the BPS conditions}

Given the backgrounds described above, we proceed with the analysis of the
conditions for unbroken supersymmetry. This is parallel to the discussion in
section \ref{sec-stat-attr}, but differs in that we only analyze the supersymmetry
preserved by the attractors corresponding to asymptotically flat black holes as given by \eqref{rot-attr-metr}
rather than derive the general conditions for $1/4$-BPS backgrounds. This
is because there is at present no evidence that asymptotically AdS
under-rotating black holes can be constructed and the near horizon
properties of such hypothetical solutions is unclear. However, we note that there is no argument against the existence of such solutions in AdS and one can try to generalize our analysis by rescaling the sizes of the AdS$_2$ and S$^2$ also in the rotating case.

We now turn to the analysis, starting with the gravitino variation and imposing
\eqref{6:KS-ansatz} and \eqref{KS-time} on the spinor $\epsilon_A$.
In the conditions below, we arrange all terms with two gamma matrices
in the ${0\hat a}$ and ${23}$ components, in order to simplify calculations.
We start from the spherical components of the variation, which can be
shown to vanish if the spinor $\epsilon_A$ does not depend on $\phi$ and
the following conditions are imposed
 \begin{align}
&\,(\partial_{\theta} +\tfrac{i}2\,Q_{\theta})\varepsilon_A
+ \tfrac{i}2\, Z(\cF)^{-}_{0\theta}\, \ee^{-i\alpha} \,\varepsilon_A
=0
\,,
\\
&\,i\,\Iprod{G}{A_{\hat a}}
+i\, \omega_{\hat a}^{-\,23}
+  \ee^{-i\alpha} Z(\cF)^{-}_{0}{}^{\hat b}\,\varepsilon_{\hat b\hat a}
=0
\,,
\\
&\,\tfrac12\,\varepsilon^{\hat\mu\hat a}\, \omega_{\hat\mu}^{-\,0}{}_{\hat a} \ee^{i\alpha}
+i\, Z(\cF)^{-}_{23} +  W\,\ =0\,,
\end{align}
where the last relation represents a term present in both components.
Using \eqref{rot-attr-metr} and \eqref{gauge_fields_rot} for the metric and gauge fields, these are simplified as follows. The first leads to an equation that determines the angular dependence of the spinor as
\begin{equation}\label{theta-dep}
 2\,\partial_{\theta} \varepsilon_A
= \partial_{\theta}U\,\varepsilon_A\,,
\end{equation}
while the second relation reduces to \eqref{Dirac_quant}. Finally, the third relation boils down to
\begin{equation}\label{cond2-rot}
 Z -i\,\ee^{U}\,\mathrm{j}\,\cos\theta\,\ee^{i\alpha} + (2\,\mathrm{j}\,\cos\theta - i\, \ee^{-2U})\,W=0\,,
\end{equation}
which generalises \eqref{cond2} in the rotating case.

Turning to the AdS$_2$ part, we analyse the time component of the Killing spinor equation, which upon assuming time independence\footnote{This assumption is consistent as we eventually show that all BPS equations are satisfied and we explicitly derive the spacetime dependence of the Killing spinors, which only depend on the $r$ and $\theta$ coordinates.} of $\epsilon_A$, implies the following constraints
 \begin{align}
i\, \Iprod{G}{A_{t}}
+i\, \omega_t^{-\,23} &\,
-  r\ee^U Z(\cF)^{-}_{23}\,\ee^{-i\alpha}
- i\,r\ee^U \,W\, \ee^{-i\alpha}
=0 \,,
\\
&\, \omega_t^{-\,0}{}_{\hat a}\, \ee^{i\alpha}
= i\,r\ee^U Z(\cF)^{-}_{0\hat a}
\,,
\end{align}
where the second equation is identically satisfied by using \eqref{F-comp-rot} and \eqref{Z-zeta}.
The first relation leads to
\begin{gather}
2\, \ee^{U}\,\R (\ee^{-i\alpha}\,W) = \ee^{4U}\,\mathrm{j}\,\cos\theta \,,
\nonumber\\
 \ee^{-U} +  \ee^{-i\alpha}\, Z
+ 2\,\mathrm{j}\,\cos\theta\,  \left(\ee^{-i\alpha}\, W
-i\,\ee^{U}\,\right)
+i\,\ee^{-2U}W\, \ee^{-i\alpha} =0 \,.
\label{calibr}
\end{gather}
Finally we consider the radial component, which leads to the constraints
 \begin{gather}
 2\,\partial_r \varepsilon_A
 + r^{-1}\ee^{-U} Z(\cF)^{-}_{23}\,\ee^{-i\alpha}
+i\,r^{-1}\ee^{-U} W\,\ee^{i\alpha}  =0 \,,
\\
 \omega_r^{-\,0\hat b}\,\varepsilon_{\hat b\hat a}\, \ee^{i\alpha}
= -r^{-1}\ee^{-U} Z(\cF)^{-}_{0\hat a}\,.
\end{gather}
These are also satisfied by using \eqref{F-comp-rot} and \eqref{calibr}, for a spinor that depends on the radial coordinate according to
 \begin{align}
 \partial_r \varepsilon_A = \frac1{2\,r}\,\varepsilon_A\,,
\end{align}
where we used \eqref{cond2-rot} and \eqref{calibr} to obtain this result. Using the last equation and \eqref{theta-dep}, find that the spacetime dependence of the Killing spinors is given by
\begin{equation}
	\varepsilon_A (r,\theta) = e^{U/2}\ \sqrt{r}\  \varepsilon^0_A\ ,
\end{equation}
for arbitrary constant spinors $\varepsilon^0_A$ that obey the two projections \eqref{6:KS-ansatz} and \eqref{KS-time} imposed above.

In addition, we need to consider the BPS conditions arising from the gaugino variation in \eqref{susy-vars}, which in this case lead to
\begin{align}
 \ee^{2U}\,Z_i + 2\,\ee^{2U}\,\mathrm{j} \cos\theta\,Z_i(\zeta) -i\,W_{i}=0\,,
\qquad
\partial_{\hat a} t^i = i\,\ee^{i\alpha}\, \bar Z_i(\partial_{\hat a}\zeta)
\,.
\end{align}
The second condition is identically satisfied upon using the $\zeta$ given in
\eqref{gauge_fields_rot}, whereas the first reads
\begin{equation}\label{gaugino-rot}
  \ee^{2U}\,Z_i + 2\,\ee^{2U}\,\mathrm{j}\,W_i \,\cos\theta -i\,W_{i}=0\,.
\end{equation}

This concludes our analysis of the BPS conditions for rotating attractors. The value of the scalar fields at the horizon can now be cast in terms of an attractor
equation generalising \eqref{univ-attr-mod} to the rotating case, as
\begin{equation}\label{univ-attr-rot}
2\,\ee^{-U}{\mathrm{Im}}\left[(1+2\,i\,\ee^{2U}\,\mathrm{j} \cos\theta)\,\ee^{-i\alpha} {\cal V}\right]
= \Gamma + \ee^{-2U}\,\mathrm{J} \cG
+ 2\,\mathrm{j} \cos\theta\, \cG\,,
\end{equation}
where one still has to impose \eqref{cond2-rot} as a constraint.

The BPS conditions above can be straightforwardly seen to be the horizon limit of the single center rotating black holes of \cite{Bossard:2012xs}, using the definition \eqref{Rstar} to simplify the result as in the static case. Since these were shown to be the most general asymptotically flat extremal under-rotating black holes, we have thus shown that all under-rotating attractor solutions are $1/4$-BPS (i.e. preserve two supercharges) when embedded in a gauged supergravity with a flat potential\footnote{The acute reader might notice that in the static case the mAdS$_2 \times$S$^2$ superalgebra $SU(1,1|1) \times SO(3)$ can be broken to $SU(1,1|1) \times U(1)$ without breaking more supersymmetries. This means that one could expect the rotating attractors to also preserve half of the original supercharges. Here we explicitly showed that these attractors are 1/4 BPS by imposing \eqref{6:KS-ansatz} and \eqref{KS-time}, but this does not exclude the existence of a more general 1/2 BPS projection that also
ensures the supersymmetry variations vanish. The flat rotating attractors here might also be part of a more general class of rotating attractors in gauged supergravity, such as the ones constructed in \cite{Klemm:2010mc}. We do not pursue this subject further as our present purpose is to show that all asymptotically flat attractors are supersymmetric without focusing on the exact amount of preserved supercharges.}. Upon taking limits of vanishing angular momentum and gaugings one finds that supersymmetry is enhanced, since the static attractors in the previous section are 1/2-BPS when $G\!\neq\!0$ and fully BPS when the gauging vanishes. In Table \ref{table-BPS} we summarise the findings of this section for all static and under-rotating attractors in abelian gauged $\cN\!=\!2$ theories.

\begin{table}[h]
 \begin{center}
\begin{tabular}{!{\vrule width 1.3pt}c|c!{\vrule width 1.3pt}c|c!{\vrule width 1.3pt}}

\Cline{1.3pt}{3-4} \multicolumn{2}{c!{\vrule width 1.3pt}}{} & $
\begin{array}{c}
\\
\mathbf{Attractor}  \\
~
\end{array}
$ & $ \begin{array}{c}
\\
 \mathbf{Global}  \\
~
\end{array}$ \\
\thickhline
\multirow{3}[3]*{  $ \begin{array}{c}
\\ \\
G \in \vsorb,\, \mathbf{flat}   \\
~
\end{array}$ } & $ \begin{array}{c}
\\
\mathrm{j}=0   \\
~
\end{array}$& $ \begin{array}{c}
\\
1/2 \  \mathrm{BPS}   \\
~
\end{array}$  & $ \begin{array}{c}
\\
\mathrm{non\!-\!BPS}   \\
~
\end{array}$ \bigstrut\\\cline{2-4}
&$ \begin{array}{c}
\\
\mathrm{j} \neq 0   \\
~
\end{array}$ & $ \begin{array}{c}
\\
1/4 \ \mathrm{BPS}   \\
~
\end{array}$ & $ \begin{array}{c}
\\
  \mathrm{non\!-\!BPS}  \\
~
\end{array}$\bigstrut\\
\Cline{1.3pt}{1-2}
 \Cline{1.3pt}{3-4}
\multirow{3}[3]*{$ \begin{array}{c}
\\ \\
G \notin \vsorb,\, \mathbf{AdS}   \\
~
\end{array}$} & $ \begin{array}{c}
\\
\mathrm{j}=0   \\
~
\end{array}$ &$ \begin{array}{c}
\\
1/2\ \mathrm{BPS}  \\
~
\end{array}$  &  $\begin{array}{c}
\\
1/4\ \mathrm{BPS} \  \\
~
\end{array}$  \bigstrut\\\cline{2-4}
& $ \begin{array}{c}
\\
\mathrm{j} \neq 0\     \\
~
\end{array}$   & $ \begin{array}{c}
\\
   ?   \\
~
\end{array}$  & $ \begin{array}{c}
\\
   ?   \\
~
\end{array}$  \bigstrut
\bigstrut\\ \thickhline

\end{tabular}
\end{center}
\caption{An overview of supersymmetry properties of under-rotating attractors and full solutions in abelian gauged theories without hypermultiplets, depending on whether the vector of gaugings $G$ lies in the very small orbit $\vsorb$ or not. The "?" for the under-rotating case in AdS signify that the existence of such solutions is not certain, not only that their supersymmetry properties are not analyzed.}\label{table-BPS}
\end{table}

\section[Asymptotically AdS BPS black holes]{Asymptotically \protect{AdS$_4$} BPS black holes}
\label{sec:AdS}

In section \ref{sec:static-ungauging} we introduced a procedure to obtain first order equations for asymptotically flat non-supersymmetric black holes by mimicking the squaring of the action that leads to asymptotically AdS$_4$ BPS black holes in an abelian gauged theory. Given the very close similarity between the equations describing the two systems, it is possible to clarify the structure of asymptotically AdS$_4$ static black holes by recycling some of the objects used in the asymptotically flat case.

In this case, the appropriate form for the metric is the one in \eqref{4dmetric}, which allows for a non-flat three dimensional base. The relevant effective action now is the one in \eqref{BPS2}, where no assumptions were made for the vector $G$. The flow equations that follow from this squaring are similar to \eqref{eq:scal-flow}-\eqref{eq:Kah-1}, with vanishing Lagrange multiplier $\varphi$, together with an equation for the nontrivial $\ee^\psi$, as in \cite{Dall'Agata:2010gj}:
\begin{eqnarray}
{\cal E} &=& 0, \label{eq:E-AdS} \\
\psi'&=&2\,e^{-U}\,{\rm Im}(e^{-i \alpha} W), \label{eq:psi-AdS}\\
Q_r + \alpha'&=& -2e^{-U}\,\,{\rm Re}(e^{-i \alpha} W)\,,
\end{eqnarray}
and we repeat the expression for ${\cal E}$,
\begin{equation}\label{E0-ads}
{\cal E} \equiv  2e^{2 \psi}\left(e^{-U}{\mathrm{Im}}(e^{-i\alpha}{\cal V})\right)'
+e^{2(\psi-U)}\, \mathrm{J} \cG
+4e^{2 \psi-U}(Q_r + \alpha'){\mathrm{Re}}(e^{-i\alpha}{\cal V})
+\Gamma\,,
\end{equation}
for the readers convenience.

Since our goal is to show the similarities between the solutions of this system to the
asymptotically flat ones, we will use an ansatz and similar definitions as in section
\ref{asym-flat-stat}. Here however, we use the same relations restricting the constant
$\mathrm{m}=0$, as one can check by analysing the asymptotic fall-off of the terms in
\eqref{E0-ads} that a nonzero $\mathrm{m}$ spoils the asymptotic behavior of the scale
factor of the metric. Thus, the role of the constant $\mathrm{m}$ is drastically
changed with respect to the asymptotically flat context, where it is "dressed" with
the Lagrange multiplier and is in fact crucial to obtain the most general static
solution.

The flow equations \eqref{E0-ads} can be simplified by defining a vector $R$ from $\cG$,
as in the non-BPS asymptotically flat case. Using the definition \eqref{Rstar} with
$\mathrm{m}=0$, we find\footnote{See \eqref{G-hat-decomp} for the general case
including $\mathrm{m}$.}
\begin{align}\label{R-def}
- |W|^2\, R =&\, \mathrm{J} \cG \,.
\end{align}
The crucial difference with the previous situation is that here $R$ is neither
constant nor small, since $\cG$ is not.
This allows to rewrite the flow equation for the section as
\begin{align}\label{E-simple}
2\ee^{2 \psi}\left(\ee^{-U}{\mathrm{Im}}(\ee^{-i\alpha}{\cal V})\right)'
-2\,\ee^{2\psi}\,|W|^2\, R
%
+4\ee^{2\psi-U}(Q_r + \alpha') {\mathrm{Re}}(\ee^{-i\alpha}{\cal V}) +\Gamma
=0
\,.
\end{align}
It order to describe solutions, me employ the natural ansatz of
\cite{Cacciatori:2009iz,Dall'Agata:2010gj, Hristov:2010ri}, which only depends on a vector
of single center harmonic functions $\cH$ as
\begin{align}\label{eq:dall-sol}
 2\,\ee^{-U}\I(\ee^{-\im\alpha}\cV)= r\,\ee^{-\psi} \cH \,,
\end{align}
and immediately leads to a vanishing K\"ahler connection, as
\begin{equation}
 Q_r + \alpha' =0\,.
\end{equation}
Note that \eqref{eq:dall-sol} reduces to the asymptotically flat solution
\eqref{scal-sol} for $\ee^\psi=r$ and $\mathrm{m}=0$, as expected.
In the more general case, equations \eqref{eq:psi-AdS} and \eqref{eq:dall-sol}
determine the function $\ee^\psi$ by
\begin{equation}\label{psi-simple}
 (\ee^\psi)' = r\,\Iprod{G}{\cH}\,,
\end{equation}
which can be easily integrated.

In order to integrate the flow equation \eqref{E-simple} above, one can follow
the direct approach of \cite{Cacciatori:2009iz,Dall'Agata:2010gj, Hristov:2010ri}, that leads to
explicit solutions (see the example below). Nevertheless, some intuition from
the asymptotically flat case can be used, in order to simplify this
process. In particular, we claim that the constraint \eqref{sympl-constr},
which we repeat here
\begin{equation} \label{sympl-constr-ads}
\frac12\, I^{'}_{4}(\cH,\cG) = \Iprod{\cG}{\cH}\, \cH
- 2 \frac{\Iprod{\cG}{\cH}^2 }{\Iprod{\cG}{R}}\, R  \, ,
\end{equation}
as written in the context of asymptotically flat solutions for very small
vectors $G$ and $R$, is relevant also in the more general case, where these
vectors are generic. Note that this might again be related to a reality
constraint on the scalar flow as in \eqref{RealT}, but we do not require
any such assumption.

In view of the similarity in the flow equations and the
fact that the ansatze in \eqref{scal-sol} and \eqref{eq:dall-sol} are related
by rescaling with a function, it is conceivable that a constraint homogeneous
in all $\cH$, $G$ and $R$ as the one in \eqref{sympl-constr-ads} may indeed
be common in the two cases. Using the explicit examples in
\cite{Cacciatori:2009iz,Dall'Agata:2010gj, Hristov:2010ri}, one can see that this is indeed
the case, as we show below.

\subsection*{Example STU solution}
In order to see how the constraint above is relevant, we consider
the STU model, defined by the prepotential
\begin{equation}\label{STU}
 F= \frac{X^1X^2X^3}{X^0}\,,
\end{equation}
as an example where fairly generic explicit solutions are known, and the expression of $R$
can be computed explicitly.
Following \cite{Cacciatori:2009iz,Dall'Agata:2010gj, Hristov:2010ri}, we choose a frame where the FI terms are
\begin{equation}
 G= \left( 0,\, g^i \, ;\, g_0,\, 0\right)^T\,,
\end{equation}
and consider a vector of single center harmonic functions
\begin{equation}\label{ads-hars}
 \cH= \left( -H^0,\, 0 \, ;\, 0,\, H_i\right)^T\,,
\end{equation}
where
\begin{equation}
 H^0 = \alpha^0 + \frac{\beta^0}{r}\,, \qquad H_i = \alpha_i + \frac{\beta_i}{r}\,.
\end{equation}
The corresponding asymptotically flat solution, where the gauging is only along the $g_0$
direction is given in Appendix \ref{app:example}. The reader can easily compare the expressions
below with those in the appendix to appreciate the close similarity of the two systems.

With the above expressions one can compute from \eqref{eq:dall-sol} that
\begin{align}\label{eq:dall-sol-Re}
 \ee^{-U}\R(\ee^{-\im\alpha}\cV)= r\,\ee^{-\psi}\,\ee^{2U}
\Big( 0,\, \tfrac12\,H^0 \,|\varepsilon^{ijk}|H_j H_k \, ;\, H_1 H_2 H_3,\, 0\Big)^T\,,
\end{align}
and
\begin{equation}
 r^{-4}\,\ee^{4\psi}\,\ee^{-4U}= 4\,H^0 H_1 H_2 H_3\,.
\end{equation}
Finally, we consider a solution to \eqref{psi-simple}, given by
\begin{equation}
 \Iprod{G}{\cH}=2\,, \qquad \ee^\psi=r^2 + c\,,
\end{equation}
where $c$ is an arbitrary constant and the first equation is a simplifying condition.

One can now impose the flow equation \eqref{E-simple} using the assumptions above,
to find the constraints
\begin{gather}
\alpha^0 g_0= \alpha^i g_i\,,\qquad
p_0 = c \alpha^0 - 2 (\beta^0)^2 g_0\,,\qquad
- q_i = c \alpha_i - 2(\beta_i)^2 g_i,
\label{floweq}
\end{gather}
where all equations are valid for each value of the index $i$ separately
and there is no implicit sum.
The explicit expression for $R$ in \eqref{R-def} then reads
\begin{equation}\label{R-STU}
 R^0 = g_0\,(H^0)^2 \,, \qquad
 R_i = g^i\,(H_i)^2\,,
\end{equation}
where again there is no implicit sum.

One can now straightforwardly evaluate the constraint \eqref{sympl-constr-ads} using the harmonic functions $\cH$ in \eqref{ads-hars} and the expression for $R$ in \eqref{R-STU}, to find that it is identically satisfied. We conclude that this constraint is also relevant for asymptotically AdS$_4$ solutions, since (similar to the asymptotically flat case) one can invert the procedure above to find $R$ from \eqref{sympl-constr-ads} rather than performing the tedious computation of the matrix $\mathrm{J}$ in \eqref{cmplx-sdual}.

Additionally, the near horizon limit of \eqref{sympl-constr-ads}, leads to a nontrivial constraint on the charges in terms of $G$, exactly as in the asymptotically flat non-BPS case. This is equivalent to the constraints found in \cite{Cacciatori:2009iz,Dall'Agata:2010gj, Hristov:2010ri} by solving the BPS flow equations in the STU model explicitly. From that point of view, \eqref{sympl-constr-ads} appears to be a duality covariant
form of the constraints on the charges in this class of solutions, valid for other symmetric models beyond STU.

\section{Extensions including hypermultiplets}
\label{sec:hypers}

Given the results of section \ref{sec:static-ungauging} on the embedding of asymptotically flat
black holes in gauged theories, it is natural to consider the possibility of extending the abelian
gauged theory to include hypermultiplets. Indeed, the appearance of the vector of gaugings $G$ multiplied
by a universal function, introduced as a Lagrange multiplier, that is determined
independently from the vector multiplet scalars, is a tantalising hint towards such an embedding.
In this scenario, one would require the gauging of a single $U(1)$ factor in the hypermultiplet
sector, where the overall Lagrange multiplier $\ee^\varphi$ in \eqref{static-rescal} is
now promoted to a dynamical field, identified with the corresponding moment map, and $G$ is identified
with the embedding tensor \cite{deWit:2005ub, deWit:2011gk}. In this section, we explore the possibilities of constructing such a theory, without explicitly considering the embedding of the known asymptotically flat solutions.

In doing so, we consider the explicit compactifications of M-theory on Calabi-Yau manifolds fibered
over a circle described in \cite{Andrianopoli:2002mf, Aharony:2008rx, Looyestijn:2010pb},
(see \cite{Cassani:2012pj} for a recent overview).
This setting is very convenient for our purposes, as it automatically leads to a flat potential for
the vector multiplets \eqref{gau-pot}, since there is only one $U(1)$ isometry gauged, along the vector
shown in \eqref{small-exam}, identified with the embedding tensor. The hypermultiplet scalars,
$q^u$, $u=1\dots 4\nh$, in these models parametrise target spaces in the image of the c-map, and describe a
fibration of a $2\nh+2$ dimensional space with coordinates $( a\,, \tilde{a}\,, \xi )$, where
$\xi=(\xi^A\,, \tilde{\xi}_{A})$ is a $2\nh$ dimensional symplectic vector, over a special K\"ahler
manifold of dimension $\nh-1$, with coordinates arranged in a complex symplectic section
$\Omega=(Z^A\,, G_{A})$, similar to the vector multiplets.

Within this setting, we consider the gauging along the Killing vector
\begin{equation}\label{gen-kill-hyp}
 k_{\mathbb{U}} \,=\, (\mathbb{U}\Omega)^A \frac{\partial}{\partial Z^A} + (\mathbb{U}\bar \Omega)^A \frac{\partial}{\partial \bar Z^A} + (\mathbb{U} \xi)^A \frac{\partial}{\partial\xi^A} + (\mathbb{U} \xi)_A \frac{\partial}{\partial\tilde\xi_A}\,,
\end{equation}
where $\mathbb{U}$ is a symplectic matrix whose explicit form can be found in \cite{Looyestijn:2010pb, Cassani:2012pj}, but is not of immediate importance for what follows. This leads to the standard minimal coupling
term for hypermultiplets, by replacing derivatives by covariant derivatives in the kinetic term as
\begin{equation}
h_{uv}\, D_\mu q^u D^\mu q^v
 = h_{uv}\big(\partial_\mu q^u + k_{\mathbb{U}}^u\, \Iprod{G}{A_\mu}\big)
    \big(\partial^\mu q^v + k_{\mathbb{U}}^v\, \Iprod{G}{A^\mu}\big)\,,
\end{equation}
where $h_{uv}$ denotes the hyper-K\"ahler metric. The potential of the gauged theory
is now given by
\begin{equation}\label{V-total}
 V = V_g + V_{\mathrm{hyp}}\,,
\end{equation}
where $V_g$ is a modification of \eqref{gau-pot} as
\begin{equation}\label{gau-pot-hyp}
V_g= Z_i(\cG)\,\bar Z^i(\cG)\,{\cal P}^2 -3\,\left|Z(\cG)\right|^2\,{\cal P}^2
= \Iprod{\cG}{\mathrm{J}\,\cG}\,{\cal P}^2 -4 \left|Z(\cG)\right|^2\,{\cal P}^2 \,,
\end{equation}
and $\,{\cal P}^2$ stands for the square of the triplet of moment maps,
$\{P^x,\,P^y,\,P^z\}$ corresponding to \eqref{gen-kill-hyp}, given by
\begin{equation}\label{P-def}
 {\cal P}^2 = (P^x)^2 + (P^y)^2+(P^z)^2\,.
\end{equation}
The second term in the scalar potential \eqref{V-total} arises from the hypermultiplet
gauging and reads
\begin{equation}
 V_{\mathrm{hyp}}= 8\,h_{uv}\,k_{\mathbb{U}}^u\, k_{\mathbb{U}}^v\, |\Iprod{G}{\cV}|^2\,.
\end{equation}

In order for the Einstein equation to allow for asymptotically flat solutions,
one must impose the condition
\begin{equation}\label{k2-zero}
 h_{uv}\,k_{\mathbb{U}}^u\, k_{\mathbb{U}}^v=0\,,
\end{equation}
which eliminates both the potential $V_{\mathrm{hyp}}$ and the term quadratic in gauge fields
in the scalar kinetic term.
Since the vector $G$ in \eqref{small-exam} lies in the doubly critical orbit $\vsorb$, the
vector multiplet potential \eqref{gau-pot-hyp} vanishes identically, as usual. Note however
that the vector of gaugings naturally appears multiplied by an overall
function, the moment map ${\cal P}$, coming from the hypermultiplet sector. We can further simplify \eqref{k2-zero}, using the facts that the quaternionic metric $h_{u v}$ is positive definite and only a single $U(1)$ is gauged, to find that\footnote{We thank Hagen Triendl for pointing out a mistake in a previous version of this paper.}
\begin{equation}\label{k-zero}
 k_{\mathbb{U}}^u=0\,.
\end{equation} 
This way the hypermultiplets condense to their supersymmetric constant values, a process described in detail in \cite{Hristov:2010eu}. The resulting theory is again the abelian gauged theory of section \ref{sec:static-ungauging}, since the moment map ${\cal P}$,
which also controls the gravitino gauging, is in general nonvanishing. Note that this is consistent
despite the initial presence of a charged hypermultiplet, due to the vanishing of the quadratic term in
gauge fields by \eqref{k2-zero}, which would otherwise produce a source in the Maxwell
equations.

We now show more explicitly how this can be realised in a simple example involving the
universal hypermultiplet \cite{Cecotti:1988qn}, which is present in all Type II/M-theory
compactifications to four dimensions, and is thus included in all the target spaces in the
image of the c-map described above. Moreover, the possible gaugings for this multiplet
are described by particular choices for $\mathbb{U}$ in the Killing vector
\eqref{gen-kill-hyp}. We use the following metric for the universal hypermultiplet,
parametrised by four real scalars $\rho$, $\sigma$, $\tau$ and $\chi$ as
\begin{align}\label{UHM-metric}
  {\rm d}s^2_{\mathrm{hyp}}
&= \frac 1 {\rho^2} \left( {\rm d}\rho^2 + \rho \, ({\rm d}\chi^2 + {\rm d}\tau^2) +
  \bigl({\rm d}\sigma +  \chi
    {\rm d}\tau\bigr)^2  \right)\ ,
\end{align}
which has eight Killing vectors (see appendix D of \cite{Hristov:2012bk}). Now, consider
the particular linear combination of Killing vectors
\begin{align}
  k_c = -\tau \partial_\chi + \chi \partial_\tau + \frac 12
  (\tau^2 - \chi^2) \partial_\sigma\ - c\, \partial_\sigma\ ,
\end{align}
parametrised by the arbitrary constant $c$. Now, the expression \eqref{k2-zero} becomes
\begin{align}
h_{uv}k^u_{c} k^v_{c} = \frac{\chi^2+\tau^2}{\rho} + \frac{(\chi^2+\tau^2 - 2\,c)^2}{4\rho^2} \ ,
\end{align}
which can vanish in two distinct situations. One is the physical minimum, corresponding
to \eqref{k-zero}, for which
\begin{equation}
\chi = \tau = 0\,, \quad \rho > 0\,, \quad c=  0 \,,
\label{rho-sol}
\end{equation}
while the second solution, given by
\begin{equation}
\rho = - \frac{(\chi^2+\tau^2 - 2\,c)^2}{4 (\chi^2+\tau^2)} < 0 \,,
\label{rho-sol-unp}
\end{equation}
is unphysical, since the metric \eqref{UHM-metric} is only positive definite
when $0<\rho<\infty$ and \eqref{rho-sol-unp} implies that it is of signature 
$(2,2)$ instead.

The triplet of moment maps associated to the Killing vector above is given by
\begin{align}
  P = \left\{\frac \tau {\sqrt{|\rho|}}\,, \,\frac \chi {\sqrt{|\rho|}}\,, \, 1 -
   \frac{\chi^2+\tau^2 -2\,c}{4\rho} \right\}\ , 
\end{align}
and reduces to $P=\{0,0,1\}$ for the physical solution in \eqref{rho-sol}, while
it is a nontrivial function of $\chi$ and $\tau$ when pulled back on the hypersurface
defined by \eqref{rho-sol-unp}.
This is an explicit realisation of the scenario sketched above, since we have obtained
a gauged theory with an everywhere vanishing scalar potential, but with nontrivial moment
maps. From this standpoint, the embedding of the asymptotically flat solution of section
\ref{sec:static-ungauging} applies directly to these models, similar to examples
discussed in \cite{Hristov:2010eu}.

From this simple discussion it follows that a single $U(1)$ hypermultiplet
gauging can only lead to asymptotically flat solutions with the hypermultiplet
scalars fixed to constants, leading to a constant moment map. If the resulting
value of the moment map is non-zero we are back in the case of FI term gauging
that allows for a BPS horizon but non-BPS asymptotics at infinity. If on the other
hand the moment maps vanish, one is in the ungauged case with BPS Minkowski vacuum
and non-BPS horizon. Note however, that the existence of unphysical solutions of
the type \eqref{rho-sol-unp} may be helpful in constructing new solutions without
hypermultiplets, where the remaining unfixed scalars may play the role of the
unphysical Lagrange multiplier in section \ref{sec:static-ungauging}.

Finally, the interesting problem of obtaining solutions with physical charged
hypermultiplets remains. Given the above, it is clear that one must gauge bigger
groups (at least $U(1)^2$) of the hypermultiplet isometries to construct such
BPS solutions, preserving supersymmetry both at the horizon and at infinity,
see e.g.\ \cite{Louis:2009xd, Louis:2012ux}.
We can then expect that such a theory, if existent, would allow for solutions
preserving two supercharges in the bulk ($1/4$-BPS solution), given that the 
attractor preserves $\cN\!=\!1$ supersymmetry, as established in section
\ref{sec:BPShor}. Constructing such a theory seems to be a nontrivial but
rather interesting task for future investigations.

\section{Conclusion and outlook}
\label{sec:conc}

In this paper we presented in some detail a novel connection between the solutions
of ungauged supergravity and gauged supergravity with an identically flat potential
in four dimensions. In particular, we identified the recently constructed general
solution for under-rotating asymptotically flat black holes as special solutions
to abelian gauged supergravity with a flat potential, where nontrivial gaugings
are still present and are reflected on the solutions. As an application, we
further showed explicitly that the attractor geometries of these black holes
belong to the generic class of 1/2-BPS AdS$_2\times$S$^2$ attractor backgrounds
that pertain to (generically asymptotically AdS$_4$) black holes in abelian gauged
theories. These results are interesting from several points of view, respectively
discussed in the main sections above. In this final section, we discuss the
implications of our results for possible string models of extremal black holes
as well some intriguing similarities to recent results in the study of black
holes in the context of supergravity.

The somewhat surprising result of obtaining hitherto hidden supercharges in the
near horizon geometries of all extremal under-rotating black holes, deserves some
additional attention. Firstly, the supercharges at hand only exist when appropriate
FI terms are turned on for given charges, and are not present in
general. This means that not all attractors characterised by a negative quartic
invariant $I_4(\Gamma)$ of the charges can be made supersymmetric simultaneously,
in contrast to the ones with $I_4(\Gamma)\!>\!0$, which correspond to globally
BPS solutions. This situation is reminiscent of the example solutions studied
recently in \cite{Bena:2011pi, Bena:2012ub}, which preserve supersymmetry only
when embedded in a larger theory, but appear as non-BPS in any $\cN\!=\!2$
truncation. In combination with these examples, our results show that any supercharges
preserved by a solution in a higher dimensional theory, may only be realised in
more general compactifications as opposed to a naive dimensional reduction.
This fits very well with the fact that asymptotically
Taub-NUT non-BPS black holes in five dimensions can preserve the full
$\cN\!=\!2$ supersymmetry near their horizons \cite{Goldstein:2008fq}. Indeed,
while a direct dimensional reduction along the Taub-NUT fiber breaks all
supersymmetries, our results imply that the Scherk-Schwarz reduction of
\cite{Looyestijn:2010pb} preserves half of them. It would be interesting to
develop a higher dimensional description of our alternative embedding along
these lines, especially in connection to possible microscopic models.

Indeed, one of the most intriguing implications of the presence of supersymmetry
for asymptotically flat under-rotating black holes is the possibility of obtaining
control over the microscopic counting of the entropy, similar to globally
BPS black holes. According to standard lore, one expects a dual microscopic CFT
living on the worldvolume of appropriate D-branes to be the relevant description
at weak coupling. Such models have been proposed in
e.g.~\cite{Emparan:2006it, Emparan:2007en, Reall:2007jv, Horowitz:2007xq, Gimon:2009gk},
and arguments on the extrapolation of the entropy counting for non-BPS black holes were
formulated in \cite{Dabholkar:2006tb, Astefanesei:2006sy}, based on
extremality. However, our results indicate that one may be able to do better, if a
supersymmetric CFT dual for extremal black holes can be found. At this point, one is tempted
to conjecture that such a CFT should be a deformation of the known theories describing
BPS black holes \cite{Strominger:1996sh, Maldacena:1997de}, where half of the
supercharges are broken by the presence of appropriate deformation parameters,
corresponding to nonzero FI terms. Obtaining a description along
these lines would be also very interesting from the point of view of black hole
physics in AdS, since it would shed some light on the role of the gaugings in a
microscopic setting.

A related question in this respect is the possibility of extending our embedding of
asymptotically flat solutions to theories including gauged hypermultiplets. As briefly
discussed in section \ref{sec:hypers}, such models with identically flat potentials are
possible, and exploring the various gaugings allowed is an interesting subject on its
own. From a higher dimensional point of view, the particular gaugings described in
\cite{Looyestijn:2010pb} represent a natural choice, since they can be formulated in
terms of a twisted reduction of ungauged five dimensional supergravity along a circle.
A similar twist was recently used in \cite{Bena:2012wc}
in connection to the near horizon geometry of over-rotating black holes.

Parallel to the implications on the asymptotically flat solutions, one may use the
connection established here to learn more about 1/4-BPS black holes in AdS. The somehow
surprising fact that the constraint on the charges defined in \cite{Bossard:2012xs}
in the case of flat gauging, is relevant in the more general setting where the gaugings
are unrestricted is a hint towards a better understanding of the moduli space of these
solutions. Indeed, since this constraint is relevant throughout the flow connecting two
uniquely fixed vacua, the asymptotic AdS$_4$ vacuum of the theory and the BPS attractor,
it may be relevant for establishing existence criteria for given charges.
In addition, it would be interesting to extend our procedure to the non-extremal case,
by connecting the results of \cite{Klemm:2012yg,Toldo:2012ec} with those of
\cite{Galli:2011fq}. We hope to return to some of these questions in future work.

\section*{Acknowledgement}
We thank Guillaume Bossard and Gianguido Dall'Agata for fruitful discussions and
useful comments on an earlier draft of this paper. We further acknowledge helpful
discussions on various aspects of this work with Iosif Bena, Bernard de Wit,
Kevin Goldstein, Hagen Triendl and Stefan Vandoren. K.H. is supported in part
by the MIUR-FIRB grant RBFR10QS5J "String Theory and Fundamental Interactions".
S.K. and V.P are supported by the French ANR contract 05-BLAN-NT09-573739, the
ERC Advanced Grant no. 226371, the ITN programme PITN-GA-2009-237920 and
the IFCPAR programme 4104-2.


\begin{appendix}

\section[Conventions on N=2 supergravity]{Conventions on $\cN\!=\!2$ supergravity}
\label{app:conv}

In this paper we follow the notation and conventions of \cite{Bossard:2012xs}.
In this appendix we collect some basic definitions that are useful in the main
text, referring to that paper for more details.

The vector fields naturally arrange in a symplectic vector of electric and magnetic
gauge field strengths, whose integral over a sphere defines the associated
electromagnetic charges as
\begin{equation}\label{eq:dual-gauge}
 \cF_{\mu\nu}=\begin{pmatrix} F_{\mu\nu}^I\\ G_{I\, \mu\nu}\end{pmatrix}\,,
\qquad
\Gamma=\begin{pmatrix} p^I\\ q_{I}\end{pmatrix}
 =\frac{1}{2\pi} \,\int_{S^2} \cF\,.
\end{equation}

The physical scalar fields $t^i$, which parametrize a special K\"ahler space of complex dimension $\nv$, appear through the so called symplectic section, $\cV$.
Choosing a basis, this section can be written in components in terms of scalars
$X^I$ as
\begin{equation}\label{eq:sym-sec}
\cV=\begin{pmatrix} X^I\\ F_I\end{pmatrix}\,, \qquad
F_I= \frac{\partial F}{\partial X^I}\,,
\end{equation}
where $F$ is a holomorphic function of degree two, called the prepotential,
which we will always consider to be cubic
\begin{equation}\label{prep-def}
F=-\frac{1}{6}c_{ijk}\frac{X^i X^j X^k}{X^0} \,,
\end{equation}
for completely symmetric $c_{ijk}$, $i=1,\dots \nv$. The section $\mathcal{V}$ is subject to the constraints
\begin{equation}
  \label{eq:D-gauge}
  \Iprod{\bar{\mathcal{V}}}{\mathcal{V}} =  i \qquad
  \Iprod{\bar{D}_{\bar i}\bar{\mathcal{V}}}{D_j\mathcal{V}} =  -i\,g_{\bar{i}j} \,,
\end{equation}
with all other inner products vanishing, and is uniquely determined by the physical scalar fields $t^i=\frac{X^i}{X^0}$ up to a local $U(1)$ transformation. Here, $g_{\bar{\imath}j}$ is the K\"ahler metric and the K\"ahler covariant derivative $D_i\cV$ contains the K\"{a}hler connection $Q_\mu$, defined through the K\"{a}hler potential as
\begin{equation} Q = \I[ \partial_ i \cK\, dt^i] \label{Kah-def}\ ,\qquad
 \cK = - \mbox{ln}\left( \tfrac{i}6\, c_{ijk} (t-\bar t)^i(t-\bar t)^j(t-\bar t)^k\right) \,. \end{equation}

We introduce the following notation for any symplectic vector $\Gamma$
\begin{align}\label{ch-def}
Z(\Gamma) = \Iprod{\Gamma}{\cV} \,,\\
Z_i(\Gamma) = \Iprod{\Gamma}{D_i \cV} \,,
\end{align}
with the understanding that when an argument does not appear explicitly, the
vector of charges in \eqref{eq:dual-gauge} should be inserted. In addition,
when the argument is form valued, the operation is applied component wise.
With these definitions it is possible to introduce a scalar dependent complex
basis for symplectic vectors, given by $(\cV,\, D_i\cV)$, so that any vector
$\Gamma$ can be expanded as
\begin{equation}\label{Z-expand}
\Gamma = 2 \I[- \bar{Z}(\Gamma)\,\cV + g^{\bar \imath j} \bar{Z}_{\bar \imath}(\Gamma)\, D_j \cV]\,,
\end{equation}
whereas the symplectic inner product can be expressed as
\begin{equation}\label{inter-prod-Z}
\Iprod{\Gamma_1}{\Gamma_2} = 2 \I[- Z(\Gamma_1)\,\bar{Z}(\Gamma_2)
   +g^{i\bar \jmath} Z_i(\Gamma_1) \, \bar{Z}_{\bar \jmath}(\Gamma_2)]\,.
\end{equation}
In addition, we introduce the scalar dependent complex structure $\mathrm{J}$,
defined as
\begin{equation}\label{CY-hodge}
\mathrm{J}\cV=-i \cV\,,\quad
\mathrm{J} D_i\cV=i  D_i\cV\,,
\end{equation}
which can be solved to determine $\mathrm{J}$ in terms of the period matrix $\cN_{IJ}$
in \eqref{G-def}, see e.g.~\cite{Ceresole:1995ca} for more details.
With this definition, we can express the complex self-duality of the gauge field strengths as
\begin{equation}
 \mathrm{J}\,\cF=-*\cF\,,\label{cmplx-sdual}
\end{equation}
which is the duality covariant form of the relation between electric and magnetic
components. Finally, we record the important relation
\begin{equation}\label{VBH-def}
 \Iprod{\Gamma}{\mathrm{J}\,\Gamma}=
 |Z(\Gamma)|^2 + g^{i\bar \jmath} Z_i(\Gamma) \, \bar{Z}_{\bar \jmath}(\Gamma)
 \equiv V_{\text{\tiny BH}}(\Gamma)\,,
\end{equation}
where we defined the black hole potential $V_{\text{\tiny BH}}(\Gamma)$.

\section{First order flows for rotating black holes}
\label{app:rot}

In this appendix we discuss the rewriting of the effective action as a sum of squares
and the corresponding flow equations for stationary black holes in four dimensional
abelian gauged $\cN=2$ supergravity. In section \ref{gau-sq-rot} we present the general
case, while in section \ref{flat-rot} we specialise to the case of flat potential
to show that the general asymptotically flat under-rotating black holes are indeed
solutions of the theory in this limit.
We largely follow \cite{Denef:2000nb, Dall'Agata:2010gj} with respect to the
method and notational conventions.

\subsection{Squaring of the action}
\label{gau-sq-rot}

We start with a timelike reduction to three spatial dimensions using the metric ansatz:
\begin{gather}\label{4dmetric-rot}
  ds^2 = -e^{2U}(d t+\omega)^2
 + e^{-2U} \left( dr^2 + \ee^{2\psi} d\theta^2 + \ee^{2\psi} \sin^2{\theta} d\phi^2 \right)\,,
\end{gather}
which generalises \eqref{4dmetric} by the addition of the angular momentum vector $\omega$.
In this setting we allow for a dependence of the fields on all spatial coordinates, so
that a timelike reduction is appropriate \cite{Breitenlohner:1987dg}. The effective
three-dimensional action reads:
\begin{align}\label{red-lagr}
S_{\text{4D}}=-\frac{1}{16\pi}\int d t \int_{\mathbb{M}^3}\Big[&\,
 -2\, \left(d  \psi\wedge \star d  \psi -\star 1\right) + 2 d  U\wedge\star d U - \tfrac{1}{2}\ee^{4U} d \omega\wedge\star d \omega
\nonumber\\ \Big. & +2g_{i\bar{\jmath}}\, d t^i\wedge\star d \bar{t}^{\bar{\jmath}} +(\cF,\cF)
+ e^{-2U}\Iprod{\cG}{\diamond \cG}\star 1 -8  e^{-2U}\left|Z(\cG)\right|^2\star 1 \Big]\,,
\end{align}
where $\cF$ is the spatial projection of the four dimensional field strengths, $\star$
denotes the Hodge dual in three dimensions and we discarded a boundary term. The scalar
dependent inner product denoted by $(,)$ is a generalisation of \eqref{VBH-def} in the
rotating case that is explicitly given by \cite{Denef:2000nb}
 \begin{equation}
 \label{Prod2forms}
 (\mathcal{A},\mathcal{B}) = \frac{\ee^{2U}}{1-w^2} \int\!\mathcal{A}\wedge\big[
 \star(\mathrm{J}\,\mathcal{B}) -
 \star(w\wedge\mathrm{J}\,\mathcal{B})\,w
 +\star(w\wedge\star\mathcal{B})\big]\,,
\end{equation}
for any two symplectic vectors of two-forms $\mathcal{A}$, $\mathcal{B}$, and we define $w=\ee^{2U}\omega$, as a shorthand below.
In order to stay as close as possible to the static case, we treat the gauging parameters
$\cG$ as gauge field strengths in the three dimensional base space. We then define
\begin{equation}\label{G-def-form}
\cbG= \cG\,\star\eta\,.
\end{equation}
where $\eta$ is a one-form which we require to be invariant under the vector $\omega$, but is
otherwise undetermined at this stage. The choice $\eta=dr$ is the one relevant for the static
solutions.

Inspired by \cite{Dall'Agata:2010gj}, we can use the above definitions to recombine the gauge kinetic term and the potential using the following combination ($Z(\cbG)=\Iprod{\cbG}{\cV}$)
\begin{equation}\label{eq:tild-F}
\tilde\cF=\cF- e^{-2U}\left(\mathrm{J} \cG -w\wedge\star \cbG\right)
+4\, e^{-2U}\R\left[\R Z(\ee^{-\im\alpha}\cbG) \ee^{\im\alpha}\cV
  + i \R Z(\ee^{-\im\alpha}\star\cbG)\wedge w \ee^{\im\alpha}\cV\right]\,,
\end{equation}
which is such that
\begin{align}
(\tilde\cF, \tilde\cF)=& (\cF,\cF) +
e^{-2U}\Iprod{\cG}{\mathrm{J} \cG}\star \eta\wedge\eta
+2 \Iprod{\cF}{\star\cbG}
+8 \I\ee^{\im\alpha} Z(\cF) \wedge \R\ee^{-\im\alpha} Z(\star\cbG)
\,,
\end{align}
and $\ee^{\im\alpha}$ is an arbitrary phase as in the static case.
The scalars can be repackaged in a similar way using the standard combination \cite{Denef:2000nb}
\begin{gather}\label{W-scal}
\Ww=2\I\star D(\ee^{-U}\ee^{-i\alpha} \cV)
  -2\R D(\ee^{U}\ee^{-i\alpha}\cV\,\omega)\,,\\
 D= d+i\,( Q+ d\alpha+\tfrac{1}{2}\ee^{2U}\star d\omega)\,,
\end{gather}
which in turn is such that
\begin{gather}
2\, d U\wedge\star d U - \tfrac{1}{2}\ee^{4U} d \omega\wedge\star d \omega
+2\,g_{i\bar{\jmath}}\, d t^i\wedge\star d \bar{t}^{\bar{\jmath}}\nonumber\\
=(\Ww,\Ww) -2\,( Q+ d\alpha +\tfrac{1}{2}\ee^{2U}
\star d\omega)\wedge d w + d\,[\,2w\wedge( Q+ d\alpha)\,]\,,
\end{gather}
so that the action reads
\begin{align}
S_{\text{4D}}=-\frac{1}{16\pi}\int\!d^4 x\Big[&\, (\Ww,\Ww)
-2\,( Q+ d\alpha+ \tfrac{1}{2}\ee^{2U}
\star d\omega)\wedge d w \nonumber\\
&+(\tilde\cF,\tilde\cF)
- 2 \Iprod{\cF}{\star\cbG}
-8  e^{-2U}\left|Z(\cG)\right|^2\star 1\nonumber\\
&-8 \Big(\I(\ee^{-\im\alpha} Z(\cF)) -w\wedge \I(\ee^{\im\alpha} Z(\star\cbG))\Big)\wedge  \R\ee^{-\im\alpha} Z(\star\cbG)
\nonumber\\ \Big. & -2\,  \left( d \psi\wedge \star d  \psi -\star 1\right)
 \Big]\,.
\end{align}

We now proceed to write the action as a sum of squares, making use of the further definitions
\begin{gather}
\cE= \tilde\cF -\Ww\,,\\
\I\langle\cE,\ee^U \ee^{-\im\alpha}\cV\rangle=
e^{U}\I \left(\ee^{-\im\alpha}Z(\cF)\right)+ e^{-U}\R\left( \ee^{-\im\alpha}Z(\cbG) \right)
 - e^{-U} w\wedge \I \left(\ee^{-\im\alpha}Z(\star\cbG)\right)
+\tfrac12  d w\,.
\end{gather}
After some rearrangements one obtains the result
\begin{align}
S_{\text{4D}}
=-\frac{1}{16\pi}\int d t \int_{\mathbb{R}^3}\Big[&\, (\cE,\cE) -4\,( Q+ d \alpha +2 e^{-U}\R Z(\star\cbG) +\tfrac{1}{2}\ee^{2U}\boldsymbol{\star} d \omega) \wedge\I\langle\cE,\ee^U \ee^{-\im\alpha}\cV\rangle \nonumber\\
& - 2 \left[\Iprod{\cF+2\,\R d(\ee^{U}\ee^{-\im\alpha}\cV\,\omega)}{\star\cbG}-\star 1\right]
\nonumber\\
 &-2\, \left( \star d  \psi
-2\,e^{-U}\I (\ee^{-\im\alpha}Z(\cbG))\right)\wedge
\left(  d  \psi -2\,e^{-U}\I (\ee^{-\im\alpha}Z(\star\cbG))\right) \nonumber\\
\Big. &
+4 e^{-U}e^{2\psi}\I (\ee^{-\im\alpha}Z(\cG))\,  d \left(\ee^{-2\psi}\star\eta\right) \Big]\,.
\end{align}
Note that we added and subtracted a term in order to obtain the squaring of the third line, which leads to the
additional factor $\ee^{-2\psi}$ in the derivative of the last line.

The last form of the action is a sum of squares, except for the terms involving the derivative
of $\eta$ and $\Iprod{\cF}{\star\cbG}$, which one should demand to be a total derivative, thus
constraining the one-form $\eta$. However, since an analysis of the resulting equations of
motion is outside the scope of this appendix, we restrict ourselves to the case of an identically
flat potential, mentioning that the general equations have the same structure as the BPS
equations of \cite{Cacciatori:2008ek} and may reproduce them once $\eta$ is specified.

\subsection{Asymptotically flat solutions}
\label{flat-rot}
Turning to the asymptotically flat case, we assume that the FI terms are given by a very small
vector and we choose the one-form $\eta$ in \eqref{G-def-form} as
\begin{equation}\label{eta-flat}
 \eta=\ee^{\varphi}\,d r\,,
\end{equation}
where we absorbed the Lagrange multiplier $\varphi$ of the static squaring in section
\ref{sec:static-ungauging}, allowing it to depend on all spatial coordinates.
Similar to the static case, we impose that the base space is flat, so that $\ee^\psi=r$,
which leads to a modified rewriting of the action as
\begin{align}
S_{\text{4D}}
=-\frac{1}{16\pi}\int d t \int_{\mathbb{R}^3}\Big[&\, (\cE,\cE) -4\,( Q+ d \alpha +2 e^{-U}\R Z(\star\cbG) +\tfrac{1}{2}\ee^{2U}\boldsymbol{\star} d \omega) \wedge\I\langle\cE,\ee^U \ee^{-\im\alpha}\cV\rangle \nonumber\\
& - 2 \,\Iprod{\cF+2\,\R d(\ee^{U}\ee^{-\im\alpha}\cV\,\omega)}{\star\cbG}
+2\, d u \wedge \star d u \nonumber\\
\Big. &-2\, \left(2\,e^{-U}\I (\ee^{-\im\alpha}Z(\cbG)) -\star d u \right)\wedge
\left( 2\,e^{-U}\I (\ee^{-\im\alpha}Z(\star\cbG)) -  d u\right) \Big]\,.
\label{rot-act-mod}
\end{align}
where we discarded a non-dynamical term and $u$ is the scalar defined in \eqref{eq:Lagrange-eom}.
The equations of motion following from this action are solved by the relations
\begin{gather}
\cE= \tilde\cF -\Ww=0\,,\label{E=0}\\
 Q+ d \alpha +2 e^{-U}\R (\ee^{-\im\alpha}Z(\star\cbG))
+\tfrac{1}{2}\ee^{2U} \star d \omega=0\,,\label{Kah-conn}\\
 2\,e^{-U}\I (\ee^{-\im\alpha}Z(\star\cbG)) -  d u= 0\,,\label{d-psi-eq}
\end{gather}
along with the equation of motion for the Lagrange multiplier, which reads
\begin{equation}\label{eq:Lagrange-eom-3D}
 d \left( \star d u \right)
- d r\wedge\,\Iprod{\cG}{\cF+2\,\R d(\ee^{U}\ee^{-\im\alpha}\cV\,\omega)}\,r^{-2}\,\ee^{-2u}=0\,.
\end{equation}

Despite the apparent complication of the equations above, one can show that the rotating
black holes of \cite{Bossard:2012xs} are solutions to the equations above, in the following way.
Firstly, we introduce the decomposition of the spatial field strengths in electromagnetic
potentials and vector fields as
\begin{equation}\label{gen-gauge_fields}
 \cF=  d (\zeta\,  \omega) + d w\,,
\end{equation}
where the explicit expression for $\zeta$ follows from \eqref{eq:tild-F}, \eqref{W-scal} and \eqref{E=0}, as
\begin{equation}\label{gen-zeta}
 d \zeta= -2\,\R d(\ee^{U}\ee^{-\im\alpha}\cV ) + \star \cbG\,,
\end{equation}
whose integrability condition implies through \eqref{G-def-form} and \eqref{eta-flat} that $\eta$ is exact and thus $\ee^\varphi$ is a total derivative with respect to the radial component.
Considering a single center solution, the vector fields $dw$ define the charges $\Gamma$
through harmonic functions $\cH$, so that the equation of motion for the Lagrange multiplier
takes the form
\begin{equation}
d \left( \star d u \right)
- r^{-2}\,d r\wedge\,\Iprod{\cG}{\Gamma}\,\ee^{-2u} \star 1=0\,,
\end{equation}
and thus admits the same enveloping solution \eqref{eq:Lagrange-sol}, which we adopt henceforth. Note that this is indeed such that $\ee^\varphi$ is a total derivative as
\begin{equation}\label{mult-der}
 \ee^\varphi =\partial_r\left(\frac{1}{V}\right)\,.
\end{equation}

We are then in a position to write the linear system to be solved in the asymptotically flat case, explicitly given by
\begin{align}
 \zeta= -2\,\R (\ee^{U}\ee^{-\im\alpha}\cV ) &\, + d (\frac1{V}) \, G\,,
\label{zeta-alm}\\
 dw - \ee^{-2U}\,\hat \cbG
+4\, e^{-2U}\,\R Z(\ee^{-\im\alpha}\cbG)\,\R\left( \ee^{\im\alpha}\cV\right)
=&\,
2\I\star D(\ee^{-U}\ee^{-\im\alpha} \cV)
  -\frac1{V} \,d \omega\, G\,.
\label{rot-system}
\end{align}
This can be simplified using the definition \eqref{Rstar}-\eqref{Y-def} for the second very small
vector, and the associated decomposition
\begin{align}\label{G-hat-decomp}
\mathrm{J} \cG =&\,
-\tfrac12\, \ee^{2U}\,V^2\, R
+M\, \ee^{2U}\,\cG
+4\,M\,V\,\ee^{3U}\,
\R\Bigl(\ee^{-i\alpha}\,{\mathcal V}\Bigr)
\,,
\end{align}
that follows from it, as one can compute directly. Note that here we have upgraded the constant
$\mathrm{m}$ in \eqref{Y-def} to a function $M$, which will turn out to control the angular
momentum. Multiplying the last relation with the Lagrange multiplier in \eqref{mult-der} we find
\begin{align}
\ee^{-2U}\,\mathrm{J} \cbG =&\,
\tfrac12\,\star d V\, R
+ M\,\star d (\frac1{V}) \cG
+4\,M\,V\star d (\frac1{V})\,\ee^{U}\,
\R\Bigl(\ee^{-i\alpha}\,{\mathcal V}\Bigr)
\,,
\end{align}
which can in turn be used in \eqref{rot-system} to obtain
\begin{align}
 dw - \tfrac12\,\star d V\, R
- M\,\star d (\frac1{V}) \cG
&\, + \frac1{V} \,d \omega\, G
-2\I\star d(\ee^{-U}\ee^{-\im\alpha} {\mathcal V})
\nonumber\\
=
 &\, -4\left[2\, e^{-2U}\,\R Z(\ee^{-\im\alpha}\cbG)\,
- M\,V\star d (\frac1{V})\,\ee^{U}\,\right]
\R\Bigl(\ee^{-i\alpha}\,{\mathcal V}\Bigr)\,.\label{flow-rot-med}
\end{align}
Here and henceforth we assume that the very small vector $R$ is a constant, which will be shown to be a consistent choice at the end.
Imposing that the terms proportional to the real part of the section in \eqref{flow-rot-med} cancel, leads to the constraint
\begin{equation}\label{int-cond-rot}
 2\, e^{-2U}\,\R Z(\ee^{-\im\alpha}\cbG)= M\,V\star d (\frac1{V})\,\ee^{U}\,,
\end{equation}
which together with the additional condition
\begin{equation}\label{ang-harm}
\star d \omega = - d M\,,
\end{equation}
that implies that both $M$ and $\omega$ are harmonic, results to the system of equations
\begin{align}\label{dw-rot}
 dw - \tfrac12\,\star d V\, R
-\,\star d \left(\frac{M}{V}\right)\, \cG
=&\,
2\I\star d(\ee^{-U}\ee^{-\im\alpha} {\mathcal V})
\,.
\end{align}
Integrating the last equation leads to a generalisation of \eqref{scal-sol}, given by
\begin{align}\label{scal-sol-rot}
2\I (\ee^{-U}\ee^{-\im\alpha} {\mathcal V})
=&\, \cH - \tfrac12\,V\, R
-\,\frac{M}{V} \, \cG
\,,
\end{align}
where the vector fields are given by the harmonic functions $\cH$ as
\begin{equation}\label{dw-dh}
 d w = \star d \cH\,.
\end{equation}

One can now compare the above to the explicit equations for the general asymptotically flat under-rotating single center solutions of \cite{Bossard:2012xs}, which turn out to be described by \eqref{int-cond-rot}-\eqref{dw-rot}
with $\omega\rightarrow -\omega$ and for $R$ being a constant very small vector, as in the static case.
Indeed, one could have started from the system \eqref{rot-system} to establish that the scalar flow equations are the ones of \cite{Bossard:2012xs} up to a constraint generalising \eqref{RealT}, which is again equivalent to the constancy of the vector $R$, as we did in section \ref{sec:static-ungauging}. However, we chose to present the symplectic covariant derivation of the equations both for simplicity and completeness. The analysis of \cite{Bossard:2012xs} ensures that this is a consistent solution of the full Einstein equations, so that we do not have to consider the Hamiltonian constraint that has to be imposed on solutions to the effective action in \eqref{rot-act-mod}, as in the static case (see \cite{VanProeyen:2007pe} for details on this constraint).

\section{Example STU solution}
\label{app:example}

\subsection*{Full solution}
In this appendix we present the known rotating seed solution in a specific
duality frame \cite{Bena:2009ev}, as an
example to use in the main body. For comparison, we use the STU model, in
\eqref{STU} in what follows. The charges of the solution are given by
poles in the following choice of harmonic functions
\begin{gather}
\cH= \left( H^0,\, 0 \,; 0,\, H_i \right)^T \ ,\\
H_i= \mathrm{h}_i +\frac{q_i}{r}\,,\quad \quad H^0=\mathrm{h}^0 +\frac{p^0}{r}\,.
\end{gather}
In this duality frame the constant small vectors can be chosen as
\begin{eqnarray} \hat R=\left( -4,\, 0 \,; \, 0 ,\, 0 \right)^T\,, \qquad
    G=\left(  0 ,\, 0 \,; \, 1,\, 0\right)^T\,,  \label{SpecificRRstar} \end{eqnarray}
but we point out that the choice is not unique, see \cite{Bossard:2012xs} for a detailed discussion.
The scalar fields are given by solving \eqref{scal-sol-rot}, leading to the physical scalars
\begin{equation} t^i =\frac{ M - i e^{-2U}}{2\,H^0\,H_i}\,
\,, \label{seed-scal} \end{equation}
as well as to the real part of the section
\begin{align}\label{eq:ex-sol-Re}
 2\,\ee^{-U}\R(\ee^{-\im\alpha}\cV)= \ee^{2U}
\Big( M\,H^0,\, -H^0 \,|\varepsilon^{ijk}|H_j H_k \, ;\, -\tfrac{M^2}{H^0} -2 H_1 H_2 H_3,\, M\,H_i\Big)^T\,.
\end{align}
The metric is given by \eqref{4dmetric-rot} with
\begin{equation} e^{-4U} = 4\,H^0 H_1 H_2 H_3 - M^2 \ , \qquad \star d \omega  = - d M \,,
\label{seed-metric}\end{equation}
where $M$ is a dipole harmonic function
\begin{equation}
  M = \mathrm{m} +\frac{\mathrm{j}\,\cos\theta}{r^2}\,.
\end{equation}
Finally, the gauge fields are given by \eqref{gen-gauge_fields}, with the $\zeta$ given
by \eqref{zeta-alm} and \eqref{eq:ex-sol-Re}, while the $dw$ are given by \eqref{dw-dh}.

\subsection*{Near horizon solution}
We now take the near horizon of the solution above, which is obtained by dropping
the constants in all harmonic functions. The scalars \eqref{seed-scal} become
\begin{equation} t^i =\frac{ \mathrm{j}\,\cos\theta - i e^{-2U}}{2\,p^0\,q_i}\,
\,, \end{equation}
whereas the near horizon metric still given by
\begin{align}
  ds^2 = -e^{2U}\,r^2\,(d t+\omega)^2
 + e^{-2U} &\,\left( \frac{dr^2}{r^2} + d\theta^2 + \sin^2{\theta} d\phi^2 \right)\,,
\nonumber\\
  \ee^{-4U}= 4\,p^0\, q_1 q_2 q_3 - \mathrm{j}^2\cos^2\theta &\,
\qquad
\omega= \mathrm{j}\,\frac{\sin^2\theta}{r}\,d \phi\,.
\end{align}
For convenience we give the near horizon gauge fields, which are given by \eqref{gen-gauge_fields}
using \eqref{zeta-alm} and \eqref{dw-dh} as
\begin{align}\label{gauge_fields_hor}
 \cF= d\big[\zeta\,r\,(dt + \omega)\big] +&\, \Gamma\, \sin\theta\,d\theta\wedge d\phi\,,
\qquad
\zeta = -2\,e^{U} \R [ e^{-i \alpha } \cV ] + \tfrac1{q_0} \, G\,,
\end{align}
where the real part of the section follows from \eqref{eq:ex-sol-Re} by replacing harmonic
functions by their poles, as above.

\end{appendix}

\bibliographystyle{utphys}
\bibliography{ungauging}

\providecommand{\href}[2]{#2}\begingroup\raggedright\begin{thebibliography}{10}

\bibitem{Strominger:1996sh}
A.~Strominger and C.~Vafa, ``{Microscopic origin of the Bekenstein-Hawking
  entropy},'' \href{http://dx.doi.org/10.1016/0370-2693(96)00345-0}{{\em
  Phys.Lett.} {\bfseries B379} (1996) 99--104},
  \href{http://arxiv.org/abs/hep-th/9601029}{{\ttfamily arXiv:hep-th/9601029
  [hep-th]}}.

\bibitem{Maldacena:1997de}
J.~M. Maldacena, A.~Strominger, and E.~Witten, ``{Black hole entropy in M
  theory},'' {\em JHEP} {\bfseries 9712} (1997) 002,
  \href{http://arxiv.org/abs/hep-th/9711053}{{\ttfamily arXiv:hep-th/9711053
  [hep-th]}}.

\bibitem{Behrndt:1997ny}
K.~Behrndt, D.~Lust, and W.~A. Sabra, ``{Stationary solutions of N = 2
  supergravity},'' \href{http://dx.doi.org/10.1016/S0550-3213(97)00633-0}{{\em
  Nucl. Phys.} {\bfseries B510} (1998) 264--288},
\href{http://arxiv.org/abs/hep-th/9705169}{{\ttfamily arXiv:hep-th/9705169}}.

\bibitem{Denef:2000nb}
F.~Denef, ``{Supergravity flows and D-brane stability},'' {\em JHEP} {\bfseries
  08} (2000) 050,
\href{http://arxiv.org/abs/hep-th/0005049}{{\ttfamily arXiv:hep-th/0005049}}.

\bibitem{Gauntlett:2002nw}
J.~P. Gauntlett, J.~B. Gutowski, C.~M. Hull, S.~Pakis, and H.~S. Reall, ``{All
  supersymmetric solutions of minimal supergravity in five dimensions},''
  \href{http://dx.doi.org/10.1088/0264-9381/20/21/005}{{\em Class. Quant.
  Grav.} {\bfseries 20} (2003) 4587--4634},
\href{http://arxiv.org/abs/hep-th/0209114}{{\ttfamily arXiv:hep-th/0209114}}.

\bibitem{Khuri:1995xq}
R.~R. Khuri and T.~{Ort\'{\i}n}, ``{A Nonsupersymmetric dyonic extreme
  Reissner-Nordstrom black hole},''
  \href{http://dx.doi.org/10.1016/0370-2693(96)00139-6}{{\em Phys.Lett.}
  {\bfseries B373} (1996) 56--60},
\href{http://arxiv.org/abs/hep-th/9512178}{{\ttfamily arXiv:hep-th/9512178
  [hep-th]}}.

\bibitem{Rasheed:1995zv}
D.~Rasheed, ``{The Rotating dyonic black holes of Kaluza-Klein theory},''
  \href{http://dx.doi.org/10.1016/0550-3213(95)00396-A}{{\em Nucl. Phys.}
  {\bfseries B454} (1995) 379--401},
\href{http://arxiv.org/abs/hep-th/9505038}{{\ttfamily arXiv:hep-th/9505038}}.

\bibitem{Ortin:1996bz}
T.~{Ort\'{\i}n}, ``{Extremality versus supersymmetry in stringy black holes},''
  \href{http://dx.doi.org/10.1016/S0370-2693(98)00040-9}{{\em Phys.Lett.}
  {\bfseries B422} (1998) 93--100},
\href{http://arxiv.org/abs/hep-th/9612142}{{\ttfamily arXiv:hep-th/9612142
  [hep-th]}}.

\bibitem{Larsen:1999pp}
F.~Larsen, ``{Rotating Kaluza-Klein black holes},''
  \href{http://dx.doi.org/10.1016/S0550-3213(00)00064-X}{{\em Nucl. Phys.}
  {\bfseries B575} (2000) 211--230},
\href{http://arxiv.org/abs/hep-th/9909102}{{\ttfamily arXiv:hep-th/9909102}}.

\bibitem{LopesCardoso:2007ky}
G.~{Lopes Cardoso}, A.~Ceresole, G.~Dall'Agata, J.~M. Oberreuter, and J.~Perz,
  ``{First-order flow equations for extremal black holes in very special
  geometry},'' \href{http://dx.doi.org/10.1088/1126-6708/2007/10/063}{{\em
  JHEP} {\bfseries 10} (2007) 063},
\href{http://arxiv.org/abs/0706.3373}{{\ttfamily arXiv:0706.3373 [hep-th]}}.

\bibitem{Gimon:2007mh}
E.~G. Gimon, F.~Larsen, and J.~Simon, ``{Black Holes in Supergravity: the
  non-BPS Branch},''
  \href{http://dx.doi.org/10.1088/1126-6708/2008/01/040}{{\em JHEP} {\bfseries
  01} (2008) 040},
\href{http://arxiv.org/abs/0710.4967}{{\ttfamily arXiv:0710.4967 [hep-th]}}.

\bibitem{Goldstein:2008fq}
K.~Goldstein and S.~Katmadas, ``{Almost BPS black holes},''
  \href{http://dx.doi.org/10.1088/1126-6708/2009/05/058}{{\em JHEP} {\bfseries
  0905} (2009) 058}, \href{http://arxiv.org/abs/0812.4183}{{\ttfamily
  arXiv:0812.4183 [hep-th]}}.

\bibitem{Bossard:2009at}
G.~Bossard, H.~Nicolai, and K.~S. Stelle, ``{Universal BPS structure of
  stationary supergravity solutions},''
  \href{http://dx.doi.org/10.1088/1126-6708/2009/07/003}{{\em JHEP} {\bfseries
  07} (2009) 003},
\href{http://arxiv.org/abs/0902.4438}{{\ttfamily arXiv:0902.4438 [hep-th]}}.

\bibitem{Bena:2009ev}
I.~Bena, G.~Dall'Agata, S.~Giusto, C.~Ruef, and N.~P. Warner, ``{Non-BPS Black
  Rings and Black Holes in Taub-NUT},''
  \href{http://dx.doi.org/10.1088/1126-6708/2009/06/015}{{\em JHEP} {\bfseries
  0906} (2009) 015},
\href{http://arxiv.org/abs/0902.4526}{{\ttfamily arXiv:0902.4526 [hep-th]}}.

\bibitem{Bossard:2009my}
G.~Bossard and H.~Nicolai, ``{Multi-black holes from nilpotent Lie algebra
  orbits},'' \href{http://dx.doi.org/10.1007/s10714-009-0870-2}{{\em Gen. Rel.
  Grav.} {\bfseries 42} (2010) 509--537},
\href{http://arxiv.org/abs/0906.1987}{{\ttfamily arXiv:0906.1987 [hep-th]}}.

\bibitem{Bena:2009en}
I.~Bena, S.~Giusto, C.~Ruef, and N.~P. Warner, ``{Multi-center non-BPS black
  holes - the solution},''
  \href{http://dx.doi.org/10.1088/1126-6708/2009/11/032}{{\em JHEP} {\bfseries
  11} (2009) 032},
\href{http://arxiv.org/abs/0908.2121}{{\ttfamily arXiv:0908.2121 [hep-th]}}.

\bibitem{Bossard:2009bw}
G.~Bossard, ``{Extremal black holes and nilpotent orbits},''
\href{http://arxiv.org/abs/0910.0689}{{\ttfamily arXiv:0910.0689 [hep-th]}}.

\bibitem{Bena:2009fi}
I.~Bena, S.~Giusto, C.~Ruef, and N.~P. Warner, ``{Supergravity solutions from
  floating branes},'' \href{http://dx.doi.org/10.1007/JHEP03(2010)047}{{\em
  JHEP} {\bfseries 03} (2010) 047},
\href{http://arxiv.org/abs/0910.1860}{{\ttfamily arXiv:0910.1860 [hep-th]}}.

\bibitem{Bossard:2011kz}
G.~Bossard and C.~Ruef, ``{Interacting non-BPS black holes},''
  \href{http://dx.doi.org/10.1007/s10714-011-1256-9}{{\em Gen.Rel.Grav.}
  {\bfseries 44} (2012) 21--66},
\href{http://arxiv.org/abs/1106.5806}{{\ttfamily arXiv:1106.5806 [hep-th]}}.

\bibitem{Bossard:2012xs}
G.~Bossard and S.~Katmadas, ``{Duality covariant non-BPS first order
  systems},'' \href{http://dx.doi.org/10.1007/JHEP09(2012)100}{{\em JHEP}
  {\bfseries 1209} (2012) 100},
\href{http://arxiv.org/abs/1205.5461}{{\ttfamily arXiv:1205.5461 [hep-th]}}.

\bibitem{Ceresole:2007wx}
A.~Ceresole and G.~Dall'Agata, ``{Flow equations for non-BPS extremal black
  holes},'' {\em JHEP} {\bfseries 03} (2007) 110,
\href{http://arxiv.org/abs/hep-th/0702088}{{\ttfamily arXiv:hep-th/0702088}}.

\bibitem{Bellucci:2008sv}
S.~Bellucci, S.~Ferrara, A.~Marrani, and A.~Yeranyan, ``{STU black holes
  unveiled},''
\href{http://arxiv.org/abs/0807.3503}{{\ttfamily arXiv:0807.3503 [hep-th]}}.

\bibitem{Ceresole:2009vp}
A.~Ceresole, G.~Dall'Agata, S.~Ferrara, and A.~Yeranyan, ``{Universality of the
  superpotential for $d = 4$ extremal black holes},'' {\em Nucl.Phys.}
  {\bfseries B832} (2010) 358, \href{http://arxiv.org/abs/0910.2697}{{\ttfamily
  arXiv:0910.2697 [hep-th]}}.

\bibitem{Ceresole:2009iy}
A.~Ceresole, G.~Dall'Agata, S.~Ferrara, and A.~Yeranyan, ``{First order flows
  for ${\cal N}=2$ extremal black holes and duality invariants},''
  \href{http://dx.doi.org/10.1016/j.nuclphysb.2009.09.003}{{\em Nucl. Phys.}
  {\bfseries B824} (2010) 239--253},
\href{http://arxiv.org/abs/0908.1110}{{\ttfamily arXiv:0908.1110 [hep-th]}}.

\bibitem{Bossard:2009we}
G.~Bossard, Y.~Michel, and B.~Pioline, ``{Extremal black holes, nilpotent
  orbits and the true fake superpotential},''
  \href{http://dx.doi.org/10.1007/JHEP01(2010)038}{{\em JHEP} {\bfseries 01}
  (2010) 038},
\href{http://arxiv.org/abs/0908.1742}{{\ttfamily arXiv:0908.1742 [hep-th]}}.

\bibitem{Caldarelli:1998hg}
M.~M. Caldarelli and D.~Klemm, ``{Supersymmetry of Anti-de Sitter black
  holes},'' \href{http://dx.doi.org/10.1016/S0550-3213(98)00846-3}{{\em
  Nucl.Phys.} {\bfseries B545} (1999) 434--460},
\href{http://arxiv.org/abs/hep-th/9808097}{{\ttfamily arXiv:hep-th/9808097
  [hep-th]}}.

\bibitem{Sabra:1999ux}
W.~Sabra, ``{Anti-de Sitter BPS black holes in N=2 gauged supergravity},''
  \href{http://dx.doi.org/10.1016/S0370-2693(99)00564-X}{{\em Phys.Lett.}
  {\bfseries B458} (1999) 36--42},
  \href{http://arxiv.org/abs/hep-th/9903143}{{\ttfamily arXiv:hep-th/9903143
  [hep-th]}}.

\bibitem{Cacciatori:2009iz}
S.~L. Cacciatori and D.~Klemm, ``{Supersymmetric AdS(4) black holes and
  attractors},'' \href{http://dx.doi.org/10.1007/JHEP01(2010)085}{{\em JHEP}
  {\bfseries 1001} (2010) 085},
\href{http://arxiv.org/abs/0911.4926}{{\ttfamily arXiv:0911.4926 [hep-th]}}.

\bibitem{Dall'Agata:2010gj}
G.~Dall'Agata and A.~Gnecchi, ``{Flow equations and attractors for black holes
  in N = 2 U(1) gauged supergravity},''
  \href{http://dx.doi.org/10.1007/JHEP03(2011)037}{{\em JHEP} {\bfseries 03}
  (2011) 037},
\href{http://arxiv.org/abs/1012.3756}{{\ttfamily arXiv:1012.3756 [hep-th]}}.

\bibitem{Hristov:2010ri}
K.~Hristov and S.~Vandoren, ``{Static supersymmetric black holes in $AdS_4$
  with spherical symmetry},''
  \href{http://dx.doi.org/10.1007/JHEP04(2011)047}{{\em JHEP} {\bfseries 04}
  (2011) 047},
\href{http://arxiv.org/abs/1012.4314}{{\ttfamily arXiv:1012.4314 [hep-th]}}.

\bibitem{Klemm:2012yg}
D.~Klemm and O.~Vaughan, ``{Nonextremal black holes in gauged supergravity and
  the real formulation of special geometry},''
\href{http://arxiv.org/abs/1207.2679}{{\ttfamily arXiv:1207.2679 [hep-th]}}.

\bibitem{Toldo:2012ec}
C.~Toldo and S.~Vandoren, ``{Static nonextremal AdS4 black hole solutions},''
  \href{http://dx.doi.org/10.1007/JHEP09(2012)048}{{\em JHEP} {\bfseries 1209}
  (2012) 048},
\href{http://arxiv.org/abs/1207.3014}{{\ttfamily arXiv:1207.3014 [hep-th]}}.

\bibitem{Berkooz:2006wc}
M.~Berkooz, D.~Reichmann, and J.~Simon, ``{A Fermi Surface Model for Large
  Supersymmetric AdS(5) Black Holes},''
  \href{http://dx.doi.org/10.1088/1126-6708/2007/01/048}{{\em JHEP} {\bfseries
  0701} (2007) 048},
\href{http://arxiv.org/abs/hep-th/0604023}{{\ttfamily arXiv:hep-th/0604023
  [hep-th]}}.

\bibitem{Kinney:2005ej}
J.~Kinney, J.~M. Maldacena, S.~Minwalla, and S.~Raju, ``{An Index for 4
  dimensional super conformal theories},''
  \href{http://dx.doi.org/10.1007/s00220-007-0258-7}{{\em Commun.Math.Phys.}
  {\bfseries 275} (2007) 209--254},
\href{http://arxiv.org/abs/hep-th/0510251}{{\ttfamily arXiv:hep-th/0510251
  [hep-th]}}.

\bibitem{Bellucci:2008cb}
S.~Bellucci, S.~Ferrara, A.~Marrani, and A.~Yeranyan, ``{d=4 Black Hole
  Attractors in N=2 Supergravity with Fayet-Iliopoulos Terms},''
  \href{http://dx.doi.org/10.1103/PhysRevD.77.085027}{{\em Phys.Rev.}
  {\bfseries D77} (2008) 085027},
\href{http://arxiv.org/abs/0802.0141}{{\ttfamily arXiv:0802.0141 [hep-th]}}.

\bibitem{Kallosh:2006ib}
R.~Kallosh, N.~Sivanandam, and M.~Soroush, ``{Exact Attractive Non-BPS STU
  Black Holes},'' \href{http://dx.doi.org/10.1103/PhysRevD.74.065008}{{\em
  Phys.Rev.} {\bfseries D74} (2006) 065008},
\href{http://arxiv.org/abs/hep-th/0606263}{{\ttfamily arXiv:hep-th/0606263
  [hep-th]}}.

\bibitem{Tripathy:2005qp}
P.~K. Tripathy and S.~P. Trivedi, ``{Non-supersymmetric attractors in string
  theory},'' \href{http://dx.doi.org/10.1088/1126-6708/2006/03/022}{{\em JHEP}
  {\bfseries 0603} (2006) 022},
\href{http://arxiv.org/abs/hep-th/0511117}{{\ttfamily arXiv:hep-th/0511117
  [hep-th]}}.

\bibitem{Goldstein:2005hq}
K.~Goldstein, N.~Iizuka, R.~P. Jena, and S.~P. Trivedi, ``{Non-supersymmetric
  attractors},'' \href{http://dx.doi.org/10.1103/PhysRevD.72.124021}{{\em
  Phys.Rev.} {\bfseries D72} (2005) 124021},
\href{http://arxiv.org/abs/hep-th/0507096}{{\ttfamily arXiv:hep-th/0507096
  [hep-th]}}.

\bibitem{Sen:2005wa}
A.~Sen, ``{Black hole entropy function and the attractor mechanism in higher
  derivative gravity},''
  \href{http://dx.doi.org/10.1088/1126-6708/2005/09/038}{{\em JHEP} {\bfseries
  0509} (2005) 038},
\href{http://arxiv.org/abs/hep-th/0506177}{{\ttfamily arXiv:hep-th/0506177
  [hep-th]}}.

\bibitem{Nampuri:2007gv}
S.~Nampuri, P.~K. Tripathy, and S.~P. Trivedi, ``{On The Stability of
  Non-Supersymmetric Attractors in String Theory},''
  \href{http://dx.doi.org/10.1088/1126-6708/2007/08/054}{{\em JHEP} {\bfseries
  0708} (2007) 054},
\href{http://arxiv.org/abs/0705.4554}{{\ttfamily arXiv:0705.4554 [hep-th]}}.

\bibitem{Ferrara:2007tu}
S.~Ferrara and A.~Marrani, ``{On the Moduli Space of non-BPS Attractors for N=2
  Symmetric Manifolds},''
  \href{http://dx.doi.org/10.1016/j.physletb.2007.07.001}{{\em Phys.Lett.}
  {\bfseries B652} (2007) 111--117},
\href{http://arxiv.org/abs/0706.1667}{{\ttfamily arXiv:0706.1667 [hep-th]}}.

\bibitem{Hristov:2012bk}
K.~Hristov, ``{Lessons from the Vacuum Structure of 4d N=2 Supergravity},''
\href{http://arxiv.org/abs/1207.3830}{{\ttfamily arXiv:1207.3830 [hep-th]}}.

\bibitem{deWit:2011gk}
B.~de~Wit and M.~van Zalk, ``{Electric and magnetic charges in N=2 conformal
  supergravity theories},''
  \href{http://dx.doi.org/10.1007/JHEP10(2011)050}{{\em JHEP} {\bfseries 1110}
  (2011) 050},
\href{http://arxiv.org/abs/1107.3305}{{\ttfamily arXiv:1107.3305 [hep-th]}}.

\bibitem{Hristov:2011ye}
K.~Hristov, C.~Toldo, and S.~Vandoren, ``{On BPS bounds in D=4 N=2 gauged
  supergravity},'' \href{http://dx.doi.org/10.1007/JHEP12(2011)014}{{\em JHEP}
  {\bfseries 1112} (2011) 014},
\href{http://arxiv.org/abs/1110.2688}{{\ttfamily arXiv:1110.2688 [hep-th]}}.

\bibitem{deWit:1984pk}
B.~de~Wit and A.~Van~Proeyen, ``{Potentials and Symmetries of General Gauged
  N=2 Supergravity: Yang-Mills Models},''
\href{http://dx.doi.org/10.1016/0550-3213(84)90425-5}{{\em Nucl. Phys.}
  {\bfseries B245} (1984) 89}.

\bibitem{deWit:1984px}
B.~de~Wit, P.~G. Lauwers, and A.~Van~Proeyen, ``{Lagrangians of N=2
  Supergravity - Matter Systems},''
\href{http://dx.doi.org/10.1016/0550-3213(85)90154-3}{{\em Nucl. Phys.}
  {\bfseries B255} (1985) 569}.

\bibitem{Cremmer:1984hj}
E.~Cremmer, C.~Kounnas, A.~Van~Proeyen, J.~Derendinger, S.~Ferrara, {\em et
  al.}, ``{Vector Multiplets Coupled to N=2 Supergravity: SuperHiggs Effect,
  Flat Potentials and Geometric Structure},''
  \href{http://dx.doi.org/10.1016/0550-3213(85)90488-2}{{\em Nucl.Phys.}
  {\bfseries B250} (1985) 385}.

\bibitem{Astefanesei:2006dd}
D.~Astefanesei, K.~Goldstein, R.~P. Jena, A.~Sen, and S.~P. Trivedi,
  ``{Rotating attractors},''
  \href{http://dx.doi.org/10.1088/1126-6708/2006/10/058}{{\em JHEP} {\bfseries
  0610} (2006) 058},
\href{http://arxiv.org/abs/hep-th/0606244}{{\ttfamily arXiv:hep-th/0606244
  [hep-th]}}.

\bibitem{deWit:2005ub}
B.~de~Wit, H.~Samtleben, and M.~Trigiante, ``{Magnetic charges in local field
  theory},'' \href{http://dx.doi.org/10.1088/1126-6708/2005/09/016}{{\em JHEP}
  {\bfseries 0509} (2005) 016},
\href{http://arxiv.org/abs/hep-th/0507289}{{\ttfamily arXiv:hep-th/0507289
  [hep-th]}}.

\bibitem{Ceresole:2010nm}
A.~Ceresole, S.~Ferrara, and A.~Marrani, ``{Small N=2 Extremal Black Holes in
  Special Geometry},''
  \href{http://dx.doi.org/10.1016/j.physletb.2010.08.053}{{\em Phys.Lett.}
  {\bfseries B693} (2010) 366--372},
\href{http://arxiv.org/abs/1006.2007}{{\ttfamily arXiv:1006.2007 [hep-th]}}.

\bibitem{Borsten:2011ai}
L.~Borsten, M.~Duff, S.~Ferrara, A.~Marrani, and W.~Rubens, ``{Small Orbits},''
  \href{http://dx.doi.org/10.1103/PhysRevD.85.086002}{{\em Phys.Rev.}
  {\bfseries D85} (2012) 086002},
\href{http://arxiv.org/abs/1108.0424}{{\ttfamily arXiv:1108.0424 [hep-th]}}.

\bibitem{Galli:2010mg}
P.~Galli, K.~Goldstein, S.~Katmadas, and J.~Perz, ``{First-order flows and
  stabilisation equations for non-BPS extremal black holes},''
\href{http://arxiv.org/abs/1012.4020}{{\ttfamily arXiv:1012.4020 [hep-th]}}.

\bibitem{Bossard:2012ge}
G.~Bossard, ``{Octonionic black holes},''
  \href{http://dx.doi.org/10.1007/JHEP05(2012)113}{{\em JHEP} {\bfseries 1205}
  (2012) 113},
\href{http://arxiv.org/abs/1203.0530}{{\ttfamily arXiv:1203.0530 [hep-th]}}.

\bibitem{Andrianopoli:1996cm}
L.~Andrianopoli, M.~Bertolini, A.~Ceresole, R.~D'Auria, S.~Ferrara, {\em et
  al.}, ``{N=2 supergravity and N=2 superYang-Mills theory on general scalar
  manifolds: Symplectic covariance, gaugings and the momentum map},''
  \href{http://dx.doi.org/10.1016/S0393-0440(97)00002-8}{{\em J.Geom.Phys.}
  {\bfseries 23} (1997) 111--189},
\href{http://arxiv.org/abs/hep-th/9605032}{{\ttfamily arXiv:hep-th/9605032
  [hep-th]}}.

\bibitem{Ceresole:1995ca}
A.~Ceresole, R.~D'Auria, and S.~Ferrara, ``{The Symplectic structure of N=2
  supergravity and its central extension},''
  \href{http://dx.doi.org/10.1016/0920-5632(96)00008-4}{{\em
  Nucl.Phys.Proc.Suppl.} {\bfseries 46} (1996) 67--74},
\href{http://arxiv.org/abs/hep-th/9509160}{{\ttfamily arXiv:hep-th/9509160
  [hep-th]}}.

\bibitem{Kallosh:1993wx}
R.~Kallosh and T.~Ortin, ``{Killing spinor identities},''
\href{http://arxiv.org/abs/hep-th/9306085}{{\ttfamily arXiv:hep-th/9306085
  [hep-th]}}.

\bibitem{Hristov:2010eu}
K.~Hristov, H.~Looyestijn, and S.~Vandoren, ``{BPS black holes in N=2 D=4
  gauged supergravities},''
  \href{http://dx.doi.org/10.1007/JHEP08(2010)103}{{\em JHEP} {\bfseries 1008}
  (2010) 103},
\href{http://arxiv.org/abs/1005.3650}{{\ttfamily arXiv:1005.3650 [hep-th]}}.

\bibitem{Hristov:2009uj}
K.~Hristov, H.~Looyestijn, and S.~Vandoren, ``{Maximally supersymmetric
  solutions of D=4 N=2 gauged supergravity},''
  \href{http://dx.doi.org/10.1088/1126-6708/2009/11/115}{{\em JHEP} {\bfseries
  0911} (2009) 115},
\href{http://arxiv.org/abs/0909.1743}{{\ttfamily arXiv:0909.1743 [hep-th]}}.

\bibitem{Louis:2012ux}
J.~Louis, P.~Smyth, and H.~Triendl, ``{Supersymmetric Vacua in N=2
  Supergravity},'' \href{http://dx.doi.org/10.1007/JHEP08(2012)039}{{\em JHEP}
  {\bfseries 1208} (2012) 039},
\href{http://arxiv.org/abs/1204.3893}{{\ttfamily arXiv:1204.3893 [hep-th]}}.

\bibitem{Hristov:2011qr}
K.~Hristov, ``{On BPS Bounds in D=4 N=2 Gauged Supergravity II: General Matter
  couplings and Black Hole Masses},''
  \href{http://dx.doi.org/10.1007/JHEP03(2012)095}{{\em JHEP} {\bfseries 1203}
  (2012) 095},
\href{http://arxiv.org/abs/1112.4289}{{\ttfamily arXiv:1112.4289 [hep-th]}}.

\bibitem{Ferrara:1995ih}
S.~Ferrara, R.~Kallosh, and A.~Strominger, ``{N=2 extremal black holes},''
  \href{http://dx.doi.org/10.1103/PhysRevD.52.R5412}{{\em Phys. Rev.}
  {\bfseries D52} (1995) 5412--5416},
\href{http://arxiv.org/abs/hep-th/9508072}{{\ttfamily arXiv:hep-th/9508072}}.

\bibitem{Strominger:1996kf}
A.~Strominger, ``{Macroscopic Entropy of $N=2$ Extremal Black Holes},''
  \href{http://dx.doi.org/10.1016/0370-2693(96)00711-3}{{\em Phys. Lett.}
  {\bfseries B383} (1996) 39--43},
\href{http://arxiv.org/abs/hep-th/9602111}{{\ttfamily arXiv:hep-th/9602111}}.

\bibitem{Ferrara:1996dd}
S.~Ferrara and R.~Kallosh, ``{Supersymmetry and Attractors},''
  \href{http://dx.doi.org/10.1103/PhysRevD.54.1514}{{\em Phys. Rev.} {\bfseries
  D54} (1996) 1514--1524},
\href{http://arxiv.org/abs/hep-th/9602136}{{\ttfamily arXiv:hep-th/9602136}}.

\bibitem{Cassani:2012pj}
D.~Cassani, P.~Koerber, and O.~Varela, ``{All homogeneous N=2 M-theory
  truncations with supersymmetric AdS4 vacua},''
\href{http://arxiv.org/abs/1208.1262}{{\ttfamily arXiv:1208.1262 [hep-th]}}.

\bibitem{Klemm:2010mc}
D.~Klemm and E.~Zorzan, ``{The timelike half-supersymmetric backgrounds of N=2,
  D=4 supergravity with Fayet-Iliopoulos gauging},''
  \href{http://dx.doi.org/10.1103/PhysRevD.82.045012}{{\em Phys.Rev.}
  {\bfseries D82} (2010) 045012},
\href{http://arxiv.org/abs/1003.2974}{{\ttfamily arXiv:1003.2974 [hep-th]}}.

\bibitem{Andrianopoli:2002mf}
L.~Andrianopoli, R.~D'Auria, S.~Ferrara, and M.~Lledo, ``{Gauging of flat
  groups in four-dimensional supergravity},'' {\em JHEP} {\bfseries 0207}
  (2002) 010,
\href{http://arxiv.org/abs/hep-th/0203206}{{\ttfamily arXiv:hep-th/0203206
  [hep-th]}}.

\bibitem{Aharony:2008rx}
O.~Aharony, M.~Berkooz, J.~Louis, and A.~Micu, ``{Non-Abelian structures in
  compactifications of M-theory on seven-manifolds with SU(3) structure},''
  \href{http://dx.doi.org/10.1088/1126-6708/2008/09/108}{{\em JHEP} {\bfseries
  0809} (2008) 108},
\href{http://arxiv.org/abs/0806.1051}{{\ttfamily arXiv:0806.1051 [hep-th]}}.

\bibitem{Looyestijn:2010pb}
H.~Looyestijn, E.~Plauschinn, and S.~Vandoren, ``{New potentials from
  Scherk-Schwarz reductions},''
  \href{http://dx.doi.org/10.1007/JHEP12(2010)016}{{\em JHEP} {\bfseries 1012}
  (2010) 016},
\href{http://arxiv.org/abs/1008.4286}{{\ttfamily arXiv:1008.4286 [hep-th]}}.

\bibitem{Cecotti:1988qn}
S.~Cecotti, S.~Ferrara, and L.~Girardello, ``{Geometry of Type II Superstrings
  and the Moduli of Superconformal Field Theories},''
\href{http://dx.doi.org/10.1142/S0217751X89000972}{{\em Int.J.Mod.Phys.}
  {\bfseries A4} (1989) 2475}.

\bibitem{Louis:2009xd}
J.~Louis, P.~Smyth, and H.~Triendl, ``{Spontaneous N=2 to N=1 Supersymmetry
  Breaking in Supergravity and Type II String Theory},''
  \href{http://dx.doi.org/10.1007/JHEP02(2010)103}{{\em JHEP} {\bfseries 1002}
  (2010) 103},
\href{http://arxiv.org/abs/0911.5077}{{\ttfamily arXiv:0911.5077 [hep-th]}}.

\bibitem{Bena:2011pi}
I.~Bena, H.~Triendl, and B.~Vercnocke, ``{Camouflaged supersymmetry in
  solutions of extended supergravities},''
  \href{http://dx.doi.org/10.1103/PhysRevD.86.061701}{{\em Phys.Rev.}
  {\bfseries D86} (2012) 061701},
\href{http://arxiv.org/abs/1111.2601}{{\ttfamily arXiv:1111.2601 [hep-th]}}.

\bibitem{Bena:2012ub}
I.~Bena, H.~Triendl, and B.~Vercnocke, ``{Black Holes and Fourfolds},''
  \href{http://dx.doi.org/10.1007/JHEP08(2012)124}{{\em JHEP} {\bfseries 1208}
  (2012) 124},
\href{http://arxiv.org/abs/1206.2349}{{\ttfamily arXiv:1206.2349 [hep-th]}}.

\bibitem{Emparan:2006it}
R.~Emparan and G.~T. Horowitz, ``{Microstates of a Neutral Black Hole in M
  Theory},'' \href{http://dx.doi.org/10.1103/PhysRevLett.97.141601}{{\em
  Phys.Rev.Lett.} {\bfseries 97} (2006) 141601},
\href{http://arxiv.org/abs/hep-th/0607023}{{\ttfamily arXiv:hep-th/0607023
  [hep-th]}}.

\bibitem{Emparan:2007en}
R.~Emparan and A.~Maccarrone, ``{Statistical description of rotating
  Kaluza-Klein black holes},''
  \href{http://dx.doi.org/10.1103/PhysRevD.75.084006}{{\em Phys.Rev.}
  {\bfseries D75} (2007) 084006},
\href{http://arxiv.org/abs/hep-th/0701150}{{\ttfamily arXiv:hep-th/0701150
  [hep-th]}}.

\bibitem{Reall:2007jv}
H.~S. Reall, ``{Counting the microstates of a vacuum black ring},''
  \href{http://dx.doi.org/10.1088/1126-6708/2008/05/013}{{\em JHEP} {\bfseries
  0805} (2008) 013},
\href{http://arxiv.org/abs/0712.3226}{{\ttfamily arXiv:0712.3226 [hep-th]}}.

\bibitem{Horowitz:2007xq}
G.~T. Horowitz and M.~M. Roberts, ``{Counting the Microstates of a Kerr Black
  Hole},'' \href{http://dx.doi.org/10.1103/PhysRevLett.99.221601}{{\em
  Phys.Rev.Lett.} {\bfseries 99} (2007) 221601},
\href{http://arxiv.org/abs/0708.1346}{{\ttfamily arXiv:0708.1346 [hep-th]}}.

\bibitem{Gimon:2009gk}
E.~G. Gimon, F.~Larsen, and J.~Simon, ``{Constituent Model of Extremal non-BPS
  Black Holes},'' \href{http://dx.doi.org/10.1088/1126-6708/2009/07/052}{{\em
  JHEP} {\bfseries 0907} (2009) 052},
\href{http://arxiv.org/abs/0903.0719}{{\ttfamily arXiv:0903.0719 [hep-th]}}.

\bibitem{Dabholkar:2006tb}
A.~Dabholkar, A.~Sen, and S.~P. Trivedi, ``{Black hole microstates and
  attractor without supersymmetry},''
  \href{http://dx.doi.org/10.1088/1126-6708/2007/01/096}{{\em JHEP} {\bfseries
  0701} (2007) 096},
\href{http://arxiv.org/abs/hep-th/0611143}{{\ttfamily arXiv:hep-th/0611143
  [hep-th]}}.

\bibitem{Astefanesei:2006sy}
D.~Astefanesei, K.~Goldstein, and S.~Mahapatra, ``{Moduli and (un)attractor
  black hole thermodynamics},''
  \href{http://dx.doi.org/10.1007/s10714-008-0616-6}{{\em Gen.Rel.Grav.}
  {\bfseries 40} (2008) 2069--2105},
\href{http://arxiv.org/abs/hep-th/0611140}{{\ttfamily arXiv:hep-th/0611140
  [hep-th]}}.

\bibitem{Bena:2012wc}
I.~Bena, M.~Guica, and W.~Song, ``{Un-twisting the NHEK with spectral flows},''
\href{http://arxiv.org/abs/1203.4227}{{\ttfamily arXiv:1203.4227 [hep-th]}}.

\bibitem{Galli:2011fq}
P.~Galli, T.~{Ort\'{\i}n}, J.~Perz, and C.~S. Shahbazi, ``{Non-extremal black
  holes of ${\cal N}=2$, $d=4$ supergravity},''
  \href{http://dx.doi.org/10.1007/JHEP07(2011)041}{{\em JHEP} {\bfseries 1107}
  (2011) 041},
\href{http://arxiv.org/abs/1105.3311}{{\ttfamily arXiv:1105.3311 [hep-th]}}.

\bibitem{Breitenlohner:1987dg}
P.~Breitenlohner, D.~Maison, and G.~W. Gibbons, ``{Four-Dimensional Black Holes
  from Kaluza-Klein Theories},''
\href{http://dx.doi.org/10.1007/BF01217967}{{\em Commun. Math. Phys.}
  {\bfseries 120} (1988) 295}.

\bibitem{Cacciatori:2008ek}
S.~L. Cacciatori, D.~Klemm, D.~S. Mansi, and E.~Zorzan, ``{All timelike
  supersymmetric solutions of N=2, D=4 gauged supergravity coupled to abelian
  vector multiplets},''
  \href{http://dx.doi.org/10.1088/1126-6708/2008/05/097}{{\em JHEP} {\bfseries
  05} (2008) 097},
\href{http://arxiv.org/abs/0804.0009}{{\ttfamily arXiv:0804.0009 [hep-th]}}.

\bibitem{VanProeyen:2007pe}
A.~Van~Proeyen and B.~Vercnocke, ``{Effective action for the field equations of
  charged black holes},''
  \href{http://dx.doi.org/10.1088/0264-9381/25/3/035010}{{\em
  Class.Quant.Grav.} {\bfseries 25} (2008) 035010},
\href{http://arxiv.org/abs/0708.2829}{{\ttfamily arXiv:0708.2829 [hep-th]}}.

\end{thebibliography}\endgroup
\end{document}